%% file: bare_jrnl.tex
\documentclass[journal]{IEEEtran}

\ifCLASSINFOpdf
\else
\fi

\usepackage{times,amsmath,epsfig}
\usepackage{epstopdf}
\usepackage{color}
\usepackage{graphicx}  
\usepackage{cases}
\usepackage{algorithm}
\usepackage{algorithmic}  
\usepackage{textcomp} 
\usepackage{tikz}
\usepackage{amsthm}
\usepackage{arydshln}

\usepackage[utf8]{inputenc}
\usepackage[english]{babel}
\newtheorem{proposition}{Proposition}

\theoremstyle{plain}
  
   \ifCLASSOPTIONcompsoc 
\usepackage[caption=false,font=normalsize,labelfon
  t=sf,textfont=sf]{subfig}
  \else
\usepackage[caption=false,font=footnotesize]{subfi
  g}
  \fi

\makeatletter
  \newcommand*{\rom}[1]{\expandafter\@slowromancap\romannumeral #1@}
\makeatother

\usepackage{cite}
\usepackage{amsmath,amssymb,amsfonts}
\usepackage{algorithmic}
\usepackage{graphicx}
\usepackage{textcomp}
\usepackage{color}
\usepackage[normalem]{ulem}
\useunder{\uline}{\ul}{}
\def\BibTeX{{\rm B\kern-.05em{\sc i\kern-.025em b}\kern-.08em
    T\kern-.1667em\lower.7ex\hbox{E}\kern-.125emX}}

\newcommand\norm[1]{\left\lVert#1\right\rVert}

\usepackage{multirow}
\usepackage{algorithm}
\usepackage{algorithmic}

 \ifCLASSOPTIONcompsoc 
  \usepackage[caption=false,font=normalsize,labelfon
  t=sf,textfont=sf]{subfig}
  \else
  \usepackage[caption=false,font=footnotesize]{subfi
  g}
  \fi

\begin{document}
%
\title{3D Point Cloud Denoising via Bipartite Graph Approximation and Reweighted Graph Laplacian}
%
%
%

\author{Chinthaka~Dinesh,~\IEEEmembership{Student~Member,~IEEE,}
        Gene~Cheung,~\IEEEmembership{Senior~Member,~IEEE,}
        and~Ivan~V.~Baji\'{c},~\IEEEmembership{Senior~Member,~IEEE}
\thanks{Chinthaka Dinesh and Ivan V. Baji\'{c} are with the School of Engineering
Science, Simon Fraser University, Burnaby, BC, Canada, e-mail: hchintha@sfu.ca and ibajic@ensc.sfu.ca.}
\thanks{Gene Cheung is with the Department of Electrical Engineering \& Computer Science, York University, Toronto, Canada, e-mail: genec@yorku.ca.}
}

\maketitle

\begin{abstract}
Point cloud is a collection of 3D coordinates that are discrete geometric samples of an object's 2D surfaces. 
Imperfection in the acquisition process means that point clouds are often corrupted with noise. 
Building on recent advances in graph signal processing, we design local algorithms for 3D point cloud denoising.
Specifically, we design a reweighted graph Laplacian regularizer (RGLR) for surface normals and demonstrate its merits in rotation invariance, promotion of piecewise smoothness, and ease of optimization.
Using RGLR as a signal prior, we formulate an optimization problem with a general $\ell_p$-norm fidelity term that can explicitly model two types of independent noise: small but non-sparse noise (using $\ell_2$ fidelity term) and large but sparser noise (using $\ell_1$ fidelity term).

To establish a linear relationship between normals and 3D point coordinates, we first perform bipartite graph approximation to divide the point cloud into two disjoint node sets (red and blue). 
We then optimize the red and blue nodes' coordinates alternately.
For $\ell_2$-norm fidelity term, we iteratively solve an unconstrained quadratic programming (QP) problem, efficiently computed using conjugate gradient with a bounded condition number to ensure numerical stability. 
For $\ell_1$-norm fidelity term, we iteratively minimize an $\ell_1$-$\ell_2$ cost function sing accelerated proximal gradient (APG), where a good  step size is chosen via Lipschitz continuity analysis.
Finally, we propose simple mean and median filters for flat patches of a given point cloud to estimate the noise variance given the noise type, which in turn is used to compute a weight parameter trading off the fidelity term and signal prior in the problem formulation.
Extensive experiments show state-of-the-art denoising performance among local methods using our proposed algorithms.
\end{abstract}


%
\IEEEpeerreviewmaketitle

\vspace{-10pt}
\section{Introduction}
\label{sec:introduction}
\input{intro}

\vspace{-5pt}
\section{Related Work}
\label{sec:related}
\input{related}

\vspace{-10pt}
\section{Preliminaries}
\label{sec:prelim}
\input{prelim}

\vspace{-7pt}
\section{Signal Prior and Problem Formulation}
\label{sec:Laplacian}
\input{laplacian}

\vspace{-12pt}
\section{Optimization Algorithm}
\label{sec:optimization}
\input{optimization}


\vspace{-7pt}
\section{Weight Parameter Estimation}
\label{sec:estimate}
\input{estimate}


\vspace{-5pt}
\section{Experimental Results}
\label{sec:exp}
\input{results}

\vspace{-5pt}
\section{Conclusion}
\label{sec:conclude}
\input{conclude}

\ifCLASSOPTIONcaptionsoff
  \newpage
\fi

\bibliographystyle{IEEEtran}
\vspace{-10pt}
\bibliography{refs}

\end{document}

%% file: intro.tex
\IEEEPARstart{A}{} recent popular 3D geometric signal representation---for a wide range of image rendering applications such as augmented reality and immersive telepresence---is \textit{point cloud}~\cite{shen2013}. 
It consists of a collection of 3D points that are geometric samples of 2D surfaces of a physical object in 3D space, sometimes with attributes such as color. 
A point cloud can be acquired directly using active light sensors such as Microsoft Kinect\copyright, or induced indirectly via stereo-matching algorithms~\cite{ji2017}.
However, imperfection in the acquisition process means that the point cloud is often corrupted with non-negligible noise. 
We address the point cloud denoising problem in this paper.

Signal smoothness is commonly used as a prior for image restoration in various forms, such as high-frequency-based Tikhonov regularization,  total variation (TV), graph Laplacian regularizer, graph TV etc,~\cite{couprie2013, pang2017, chen15}.
However, two inherent difficulties make the proper definition of smoothness more challenging for point clouds.
First, though 3D points in a point cloud---unlike regularly sampled pixels in a 2D digital image---may be irregularly spaced apart, the defined notion should nonetheless promote \textit{piecewise smoothness} (PWS) in 2D surfaces, \textit{e.g.}, flat sides of a  box connected by sharp distinct transitions.  
Second, because a point cloud specifies a free object's 3D geometry, the defined smoothness notion should be \textit{rotation-invariant} in 3D space~\cite{deschaud2010}. 
These difficulties mean that image restoration methods based on conventional smoothness notions like TV \cite{couprie2013} cannot be directly applied to restore corrupted point clouds.

In this paper, leveraging on recent advances in graph spectral image processing \cite{cheung18}, we design a smoothness notion for 3D point clouds called \textit{reweighted graph Laplacian regularizer} (RGLR) defined on surface normals with three desirable properties: i) rotation invariance, ii) promotion of PWS in surface normals, and iii) ease of optimization.
We prove that RGLR is invariant to any unitary rotation matrix. 
We show that RGLR promotes PWS by analyzing its behavior for a single node pair. 
Using RGLR as a signal prior, we formulate the point cloud denoising problem as an optimization with a general $\ell_p$-norm fidelity term that can explicitly model two types of independent noise: i) non-sparse but small noise like Gaussian (using a $\ell_2$-norm fidelity term), and ii) sparser but larger noise like Laplacian (using a $\ell_1$-norm fidelity term)\footnote{Sparse large noise can also be modeled using a $\ell_0$ norm. However, $\ell_1$ norm is the closest convex approximation and leads to fast optimal algorithms.}. 

To efficiently solve the posed optimization, we first establish a linear relationship between normals and 3D point coordinates via bipartite graph approximation \cite{zeng2017} to divide the point cloud into two disjoint node sets (red and blue). 
The normal for each red node can thus be defined with respect to neighboring blue nodes' coordinates, resulting in a linear function of the red node's coordinates.
Thereafter, we optimize the red and blue nodes' coordinates alternately.
For $\ell_2$-norm fidelity term, we iteratively solve an unconstrained \textit{quadratic programming} (QP) problem, which can be efficiently computed using \textit{conjugate gradient} (CG)~\cite{shewchuk1994} with a bounded condition number via \textit{Gershgorin circle theorem}~\cite{brakken2007}, guaranteeing numerical stability.
Alternatively, the solution can be implemented as a graph filter via the \textit{Lanczos algorithm}~\cite{susnjara2015}, allowing a spectral low-pass interpretation.
For $\ell_1$-norm fidelity term, we iteratively minimize an $\ell_1$-$\ell_2$ cost function using \textit{accelerated proximal gradient} (APG)~\cite{parikh2014}, where a step size ensuring a good convergence rate is chosen based on \textit{Lipschitz continuity} analysis~\cite{combettes2005}.

Finally, we propose simple mean and median filters for flat patches of a given point cloud to estimate the noise variance given the noise type, which in turn is used to compute a weight parameter trading off the fidelity term and signal prior in the problem formulation.
Extensive experiments show state-of-the-art denoising performance among local methods using our proposed algorithms both subjectively and objectively under a number of point cloud metrics in the literature~\cite{tian2017}.

The outline of the paper is as follows.
We first overview related works in Section\;\ref{sec:related}.
Next, we define necessary concepts in Section\;\ref{sec:prelim}. 
We then present the definitions of RGLR and problem formulation in Section\;\ref{sec:Laplacian}. 
We describe our proposed optimization algorithms in Section\;\ref{sec:optimization}. 
We discuss the computation of an appropriate weight parameter and noise variance estimation in Section\;\ref{sec:estimate}. 
Finally, experimental results and conclusions are presented in Sections\;\ref{sec:exp} and\;\ref{sec:conclude} respectively.

%% file: related.tex
We divide existing related works into five categories: moving least squares (MLS)-based methods, locally optimal projection (LOP)-based methods, sparsity-based methods, graph-based method, and non-local similarity-based method.\\
\textbf{MLS-based method:} In MLS-based methods, a smooth surface is approximated from the input point cloud, and observed points are projected to the resulting surface. 
To construct the surface, \cite{alexa2003} first finds a best-fitting local reference domain for a neighborhood of points in terms of MLS.
Then a function is defined above the reference domain by fitting a polynomial function to neighboring data. 
However, if the underlying surface has high curvatures, then this method becomes unstable. 
In response, several solutions have been proposed, \textit{e.g.}, algebraic point set surfaces (APSS) \cite{guennebaud2007} and its variant \cite{guennebaud2008} and robust implicit MLS (RIMLS) \cite{oztireli2009}. 
However, these methods may over-smooth \cite{sun2015}.\\
\textbf{LOP-based methods:} LOP-based methods do not compute explicit parameters for the point cloud surface. For example, \cite{lipman2007} generates a point set that represents the underlying surface while enforcing a uniform distribution over the point cloud. 
There are two main modifications to \cite{lipman2007}. 
The first is weighted LOP (WLOP) \cite{huang2009} that provides a uniformly distributed output by preventing a given point from being too close to other neighbors. 
The second is anisotropic WLOP (AWLOP) \cite{huang2013} that preserves sharp features using an anisotropic weighting function. 
However, LOP-based methods also suffer from over-smoothing due to local operators~\cite{sun2015}.
\\
\textbf{Sparsity-based methods:} There are two main steps in sparsity-based methods. First, a sparse reconstruction of the surface normals is obtained by solving a minimization with sparsity regularization. 
Then the point positions are updated by solving another minimization based on a local planar assumption. 
Examples include \cite{mattei2017} and  \cite{sun2015} that use $\ell_1$ and $\ell_0$ regularization respectively. 
However, for large noise these methods also over-smooth or over-sharpen \cite{sun2015}.             
\\
\textbf{Non-local methods:} Non-local methods generalize \textit{non-local means} (NLM) \cite{buades2005} and BM3D \cite{dabov2007} image denoising algorithms to point cloud denoising. 
These methods rely on the self-similarity characteristic among surface patches in the point cloud. 
Methods in \cite{digne2012, deschaud2010} utilize a NLM algorithm, while a method in \cite{rosman2013} is inspired by BM3D. 
Recently, \cite{zeng2018} defines self-similarity among patches as a \textit{low-dimensional manifold model} (LDMM) \cite{osher2017}. 
Although non-local methods achieve good performance, their computational complexity is often too high. 
In contrast, our algorithms are local and do not require non-local similar patch search.
\\
\textbf{Graph-based methods:} In these methods, first, a graph (\textit{e.g.} $k$-nearest-neighbor graph) is constructed from a given point cloud model. 
In~\cite{duan2018}, a local tangent plane at each point in the constructed point cloud graph is estimated, and then each 3D point is denoised via weighted averaging of its projections on neighboring tangent planes. 
Due to averaging, sharp edges cannot be denoised properly, resulting in over-smoothing. In~\cite{schoenenberger2015}, graph total variation (GTV)-based regularization method is used for denoising, where GTV is directly applied on the 3D coordinates. 
This is not appropriate because GTV of coordinates promotes variational proximity of 3D points, and only a singular 3D point has zero GTV value. 

In our recent work, we propose a GTV-based denoising method~\cite{dinesh2018fast}, where GTV is applied on the estimated surface normals. 
However, This usage of GTV is not \textit{rotation-invariant}; \textit{i.e.}, the value of the GTV changes when the point cloud is rotated around a random axis. 

All the aforementioned methods are derived from formulations that either explicitly assume Gaussian noise, or it is unclear for what noise type the methods work well. 
Instead we propose two graph-based denoising methods explicitly for non-sparse Gaussian noise or sparse Laplacian noise. 
Our methods differ from our recent work~\cite{dinesh2018fast} as follows: i)~we design RGLR as a signal prior for point clouds with two desirable properties---rotation invariance and promotion of PWS; ii)~based on RGLR, we formulate the denoising problem as a general $\ell_p$-$\ell_2$-norm convex optimization and design two algorithms to explicitly optimize the formulated problem for the two aforementioned noise types; and iii)~we propose different algorithms to estimate noise variance for these two noise types, which is used to compute a weight parameter trading off the fidelity term and the signal prior in the formulation.

%% file: prelim.tex

\noindent\textbf{3D Point Cloud:} We define a point cloud as a set of discrete samples of 3D coordinates of an object's 2D surface in 3D space. 
Denote by $\mathbf{q}~=~\begin{bmatrix} \mathbf{q}_{1}^{T} & \hdots & \mathbf{q}_{N}^{T} \end{bmatrix}^T\in\mathbb{R}^{3N}$ the position vector for the point cloud, where $\mathbf{q}_{i}\in\mathbb{R}^{3}$ is the 3D coordinate of a point $i$ and $N$ is the number of points in the point cloud. Noise-corrupted $\mathbf{q}$ can be simply modeled as $\mathbf{q}~=~\mathbf{p}+\mathbf{e}$,
where $\mathbf{p}$ are the true 3D coordinates, $\mathbf{e}$ is a zero-mean independent and identically distributed (iid) noise, and $\mathbf{p}$, $\mathbf{e}\in\mathbb{R}^{3N}$. 
The goal is to recover $\mathbf{p}$ given only noise-currupted observation $\mathbf{q}$.
\\
\textbf{Surface Normals:} Surface normal at a point $i$ in a given 3D point cloud is a vector (denoted by $\mathbf{n}_i\in\mathbb{R}^3$) that is perpendicular to the tangent plane to that surface at $i$. 
There are numerous methods in the literature \cite{huang2001,kanatani2005,ouyang2005,gouraud1971,jin2005} to compute a normal to the surface at point $i$. 
One popular method is to fit a local plane to point $i$ and its $k$ nearest neighbors, and take the perpendicular vector to that plane \cite{huang2001,kanatani2005,ouyang2005}. 
A common alternative is to compute the normal vector as the weighted average of the normal vectors of the triangles formed by $i$ and pairs of its neighbors \cite{gouraud1971,jin2005}.
In all these methods, the surface normal at $i$ is related nonlinearly with the 3D coordinates of $i$ and its neighbors.\\ 
\textbf{Definitions in Graph Signal Processing:} We define graph signal processing (GSP) related concepts needed in our work. 
Consider an undirected graph $\mathcal{G}=(\mathcal{V}, \mathcal{E})$ composed of a node set $\mathcal{V}$ and an edge set $\mathcal{E}$ specified by $(i,j,w_{i,j})$, where $i,j\in\mathcal{V}$ and $w_{i,j}\in\mathbb{R}^{+}$ is the edge weight that reflects the similarity between nodes $i$ and $j$. 
Graph $\mathcal{G}$ can be characterized by its \textit{adjacency matrix} $\mathbf{W}$ with $\mathbf{W}(i,j)~=~\mathbf{W}(j,i)~=~w_{i,j}$. 
Further, denote by $\mathbf{D}$ the diagonal \textit{degree matrix} where entry $\mathbf{D}(i,i)=\sum_{j}w_{i,j}$. 
Given $\mathbf{W}$ and $\mathbf{D}$, the \textit{combinatorial graph Laplacian matrix}~\cite{chung1997} is defined as~$\mathbf{L}=\mathbf{D}-\mathbf{W}$.
Since $\mathbf{W}$ and $\mathbf{D}$ are real symmetric matrices, $\mathbf{L}$ is also real and symmetric. 
Moreover, $\mathbf{L}$ is a positive semi-definite (PSD) matrix; \textit{i.e.}, for any given vector $\mathbf{f}\in \mathbb{R}^{N}$, $\mathbf{f}^\top\mathbf{Lf}\geq 0$, where $N$ is the number of nodes in the graph $\mathcal{G}$. 
$\mathbf{f}$ is called a \textit{graph signal} on graph $\mathcal{G}$, where $\mathbf{f}~=~[f_{1}\hdots f_{N}]^{\top}$ and $f_i$ is a scalar value assigned to node $i$. 
One can easily show that $\mathbf{f}^\top\mathbf{Lf}$ can be rewritten as
\vspace{-10pt}
\begin{equation}
\mathbf{f}^\top\mathbf{Lf}=\sum_{i,j\in\mathcal{E}}w_{i,j}(f_i-f_j)^2,
\label{eq:GLR}
\end{equation}
which is known as \textit{graph Laplacian regularizer} (GLR)~\cite{pang2017}.\\
\textbf{Graph Construction}: In this paper, we first construct a $k$-nearest-neighbor ($k$-NN) graph to connect 3D points in a given point cloud based on Euclidean distance in 3D space \cite{wang2011}. 
Specifically, each 3D point is represented by a node and is connected through edges to its $k$ nearest neighbors, with weights that reflect inter-node similarities. 
For a given point cloud $\mathbf{p}~=~\begin{bmatrix} \mathbf{p}_{1}^{\top} & \hdots & \mathbf{p}_{N}^{\top} \end{bmatrix}^\top$, edge weight $w_{i,j}$ between nodes $i$ and $j$ is computed using the Euclidean distance between nodes $i$ and $j$ (called \textit{distance term} $w_{i,j}^p$), and the angle $\theta_{i,j}$ between $\mathbf{n}_i$ and $\mathbf{n}_j$ (called \textit{angle term} $w_{i,j}^{\theta}$):
\vspace{-5pt}
\begin{equation}
w_{i,j}=\underbrace{\exp{\left\{-\frac{|\vert\mathbf{p}_{i}-\mathbf{p}_{j}|\vert_{2}^{2}}{\sigma_{p}^{2}}\right\}}}_{w_{i,j}^p}\underbrace{\cos^{2}\theta_{i,j}}_{w_{i,j}^{\theta}},
\label{eq:edge_weight}
\vspace{-5pt}
\end{equation}
where $\sigma_{p}$ is a parameter. In words,~(\ref{eq:edge_weight}) states that edge weight $w_{i,j}$ is large (close to 1) if points $i, j$ are close and associated normal vectors are similar, and small (close to 0) otherwise.   
By the definition of inner product of $\mathbf{n}_i$ and $\mathbf{n}_j$, $\cos^{2}\theta_{i,j}$ can be written as,
\vspace{-10pt}
\begin{equation}
\cos^{2}\theta_{i,j}=(\mathbf{n}_i^\top\mathbf{n}_j)^2=\left(\frac{2-\norm{\mathbf{n}_i-\mathbf{n}_j}_2^2}{2}\right)^2.
\label{eq:angle}
\end{equation}

Given graph $\mathcal{G}$ constructed using edge weights (\ref{eq:edge_weight}), we use $\mathcal{G}$ to regularize separately each coordinate of the points' surface normals $\mathbf{N}~=~[\mathbf{n}_1^{\top};\hdots ;\mathbf{n}_{N}^{\top}]\in \mathbb{R}^{N\times 3}$; \textit{i.e.}, we consider three graph signals corresponding to $x$-, $y$- and $z$-coordinates of $\mathbf{N}$ defined as follows: $\mathbf{n}_{x}~=~[\mathbf{n}_{1,1} \hspace{5pt} \hdots \hspace{5pt} \mathbf{n}_{N,1}]^\top$, $\mathbf{n}_{y}~=~[\mathbf{n}_{1,2} \hspace{5pt} \hdots \hspace{5pt} \mathbf{n}_{N,2}]^\top$, $\mathbf{n}_{z}~=~[\mathbf{n}_{1,3} \hspace{5pt} \hdots \hspace{5pt} \mathbf{n}_{N,3}]^\top$, where $\mathbf{n}_{i,j}$ is the $j$-th entry of $\mathbf{n}_i$. 



%% file: laplacian.tex
\begin{figure}[t]
 \centering
 \includegraphics[width=1.6in]{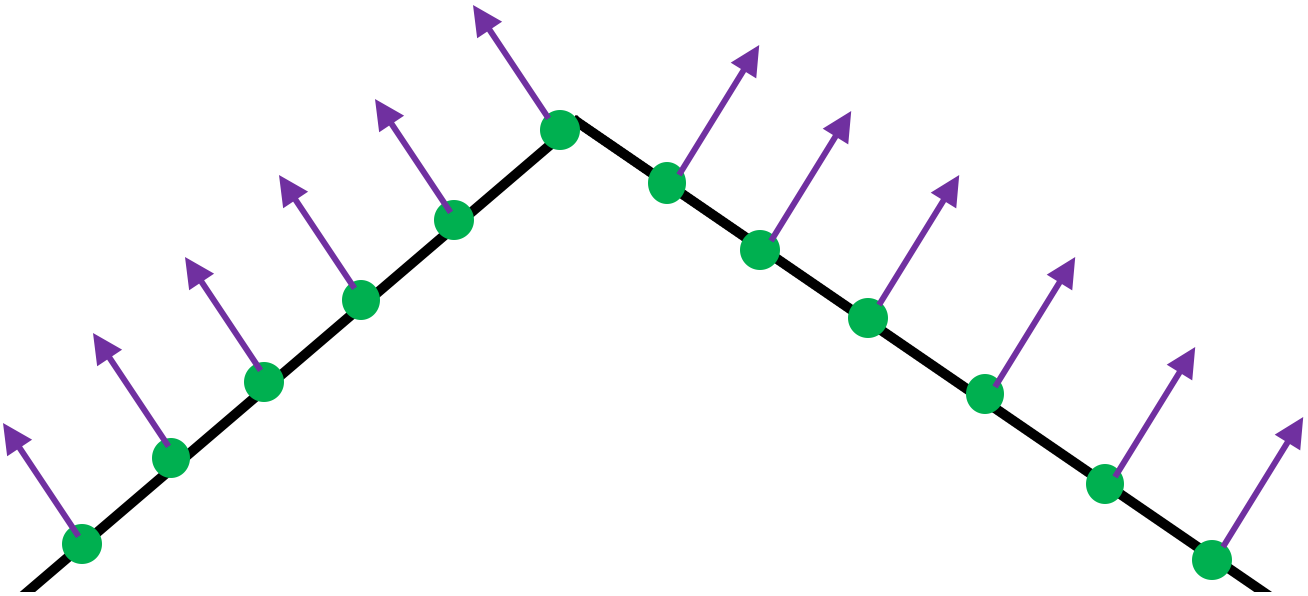} 
 \vspace{-10pt}
 \caption{A 2D demonstration of an example for PWS surface with normals.  }
 \label{fig:PWS}
\vspace{-15pt}						
  \end{figure} 
\subsection{Definition of Reweighted Graph Laplacian Regularizer}

In this paper, we assume that the normals computed (consistently oriented) at the 2D surface of a given 3D point cloud is PWS. 
In other words, given two neighboring nodes $i,j\in\mathcal{E}$ on a $k$-NN graph $\mathcal{G}~=~(\mathcal{V,E})$ representing a clean point cloud, the difference between the corresponding surface normals, \textit{i.e.} $|\vert\mathbf{n}_{i}-\mathbf{n}_{j}|\vert_2$, should be typically very small (though large occasionally). See Fig.~\ref{fig:PWS} for an illustration. 
Hence, we design a signal prior to promote PWS of $\mathbf{N}$. 
Moreover, because a point cloud specifies a free object's 3D geometry in an arbitrary 3D coordinate system, the prior should be \textit{rotation-invariant} in 3D space~\cite{deschaud2010}; \textit{i.e.}, rotating the point cloud around any axis should not change the computed prior value.    

To satisfy these two properties, we propose a RGLR prior for surface normals $\mathbf{N}$ over $\mathcal{G}$ as follows. 
Specifically, instead of using fixed $\mathbf{W}$, we extend the conventional GLR in (\ref{eq:GLR}) to RGLR, where the adjacency matrix $\mathbf{W}(\mathbf{N})$ (and hence the degree and Laplacian matrices $\mathbf{D}(\mathbf{N})$ and $\mathbf{L}(\mathbf{N})$) is a function of $\mathbf{N}$. 
We define RGLR for surface normals as 
\begin{equation}
\begin{split}
|\vert\mathbf{N}|\vert_{\text{RGLR}}&=\mathbf{n}_x^\top\mathbf{L}(\mathbf{N})\mathbf{n}_x+\mathbf{n}_y^\top\mathbf{L}(\mathbf{N})\mathbf{n}_y+\mathbf{n}_z^\top\mathbf{L}(\mathbf{N})\mathbf{n}_z\\
&=\sum_{i,j\in\mathcal{E}}w_{i,j}(\mathbf{n}_i,\mathbf{n}_j)|\vert\mathbf{n}_{i}-\mathbf{n}_{j}|\vert_2^2,
\label{eq:main_RGLR}
\end{split}
\end{equation}
where $w_{i,j}(\mathbf{n}_i,\mathbf{n}_j)$ is the $(i,j)$ element of $\mathbf{W}(\mathbf{N})$. 
The initial value of $w_{i,j}(\mathbf{n}_i,\mathbf{n}_j)$ is computed using (\ref{eq:edge_weight}). 
Here, we consider $w_{i,j}(\mathbf{n}_i,\mathbf{n}_j)$ as a function of $\mathbf{n}_i$ and $\mathbf{n}_j$ as in (\ref{eq:edge_weight}). 
We show next that our proposed RGLR in (\ref{eq:main_RGLR}) has the desired two properties: rotation invariance and promoting PWS.

\vspace{-5pt}
\begin{proposition}
RGLR defined in (\ref{eq:main_RGLR}) is invariant to any rotation matrix applied to the point cloud.
\end{proposition}
\vspace{-10pt}
\begin{proof}
Suppose we rotate a point cloud around any given axis by applying a rotational matrix\footnote{Rotation matrix is a square matrix, and a given square matrix $\mathbf{R}$ is a rotation matrix if and only if $\mathbf{R}^{\top}\mathbf{R}~=~\mathbf{I}$ and $\det \mathbf{R}~=~1$.} $\mathbf{R}\in \mathbb{R}^{3\times 3}$. 
It follows that we should apply the same $\mathbf{R}$ to all surface normals computed before the rotation. 
We can thus write RGLR in (\ref{eq:main_RGLR}) for the rotated point cloud (denoted as $\norm{\mathbf{N}}_{\text{RGLR}}^{\mathbf{R}}$) based on the surface normals computed before the rotation as follows:
\begin{equation}
\norm{\mathbf{N}}_{\text{RGLR}}^{\mathbf{R}}=\sum_{i,j\in\mathcal{E}}w_{i,j}(\mathbf{Rn}_i,\mathbf{Rn}_j)|\vert\mathbf{Rn}_{i}-\mathbf{Rn}_{j}|\vert_2^2.
\label{eq:RRGLR}
\end{equation}
According to (\ref{eq:edge_weight}) and (\ref{eq:angle}) we can write $w_{i,j}(\mathbf{Rn}_i,\mathbf{Rn}_j)$ as follows:
\vspace{-10pt}
\begin{equation}
\begin{split}
&=\exp{\left\{-\frac{|\vert\mathbf{Rp}_{i}-\mathbf{Rp}_{j}|\vert_{2}^{2}}{\sigma_{p}^{2}}\right\}}((\mathbf{Rn}_i)^\top\mathbf{Rn}_j)^2\\
&=\exp{\left\{-\frac{(\mathbf{p}_{i}-\mathbf{p}_{j})^\top\mathbf{R}^\top\mathbf{R}(\mathbf{p}_{i}-\mathbf{p}_{j})}{\sigma_{p}^{2}}\right\}}(\mathbf{n}_i^\top\mathbf{R}^{\top}\mathbf{R}\mathbf{n}_j)^2\\
&=\exp{\left\{-\frac{|\vert\mathbf{p}_{i}-\mathbf{p}_{j}|\vert_{2}^{2}}{\sigma_{p}^{2}}\right\}}(\mathbf{n}_i^\top\mathbf{n}_j)^2=w_{i,j}(\mathbf{n}_i,\mathbf{n}_j),
\end{split}
\label{eq:wrota}
\end{equation}
where $\mathbf{R}^\top\mathbf{R}=\mathbf{I}$ as $\mathbf{R}$ is a unitary matrix. 
Given (\ref{eq:wrota}), we can rewrite (\ref{eq:RRGLR}) as follows:
\vspace{-7pt}
\begin{equation}
\begin{split}
\norm{\mathbf{N}}_{\text{RGLR}}^{\mathbf{R}}&=\sum_{i,j\in\mathcal{E}}w_{i,j}(\mathbf{n}_i,\mathbf{n}_j)|\vert\mathbf{Rn}_{i}-\mathbf{Rn}_{j}|\vert_2^2\\
&=\sum_{i,j\in\mathcal{E}}w_{i,j}(\mathbf{n}_i,\mathbf{n}_j) \; (\mathbf{n}_{i}-\mathbf{n}_{j})^\top\mathbf{R}^\top\mathbf{R}(\mathbf{n}_{i}-\mathbf{n}_{j})\\
&=\sum_{i,j\in\mathcal{E}}w_{i,j}(\mathbf{n}_i,\mathbf{n}_j)|\vert\mathbf{n}_{i}-\mathbf{n}_{j}|\vert_2^2=\norm{\mathbf{N}}_{\text{RGLR}}.
\label{eq:RRGLR1}
\vspace{-10pt}
\end{split}
\end{equation}
\vspace{-0pt}
Therefore RGLR in (\ref{eq:main_RGLR}) is rotation-invariant.
\vspace{-5pt}
\end{proof}

\begin{figure}[t]
 \centering
 \includegraphics[width=3.3in]{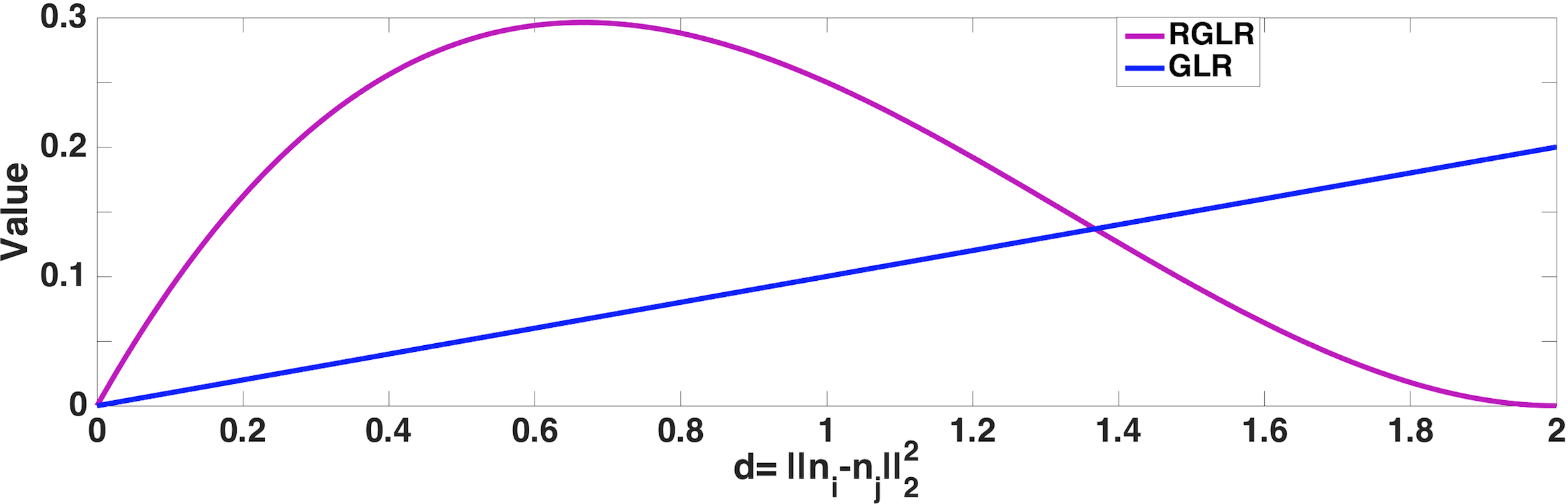} 
 \vspace{-10pt}
 \caption{Curves of RGLR and GLR for a two node graph. $w_{i,j}^p$ is set to 1 for RGLR and $w_{i,j}$ is set to 0.1 for GLR.}
 \label{fig:curves}
\vspace{-15pt}						
  \end{figure}

A simple analysis shows that the proposed RGLR promotes PWS. 
Since (\ref{eq:main_RGLR}) is a sum of separable terms, we can study the behavior of RGLR using a single node pair $(i,j)$; the behavior for the entire graph would be the sum of behaviors of individual pairs plus their interaction.
With $d~=~\norm{\mathbf{n}_i-\mathbf{n}_j}_2^2$ and fixed $w_{i,j}^p$, RGLR for pair $(i,j)$ is $w_{i,j}(\mathbf{n}_i,\mathbf{n}_j)\norm{\mathbf{n}_i-\mathbf{n}_j}_2^2~=~w_{i,j}^p((2-d)^2/4)d$. 
Here $0\leq d\leq 2$ because we consistently orient each surface normal so that $\mathbf{n}_i^T\mathbf{n}_j\geq 0$ (to be discussed in Section \ref{sec:normalvector}). 
Within this range, RGLR has one maximum at $2-\sqrt{2}$ and two minima at $0$ and $2$, as shown in Fig.\;\ref{fig:curves}. 
Hence, using gradient descent to minimize (\ref{eq:main_RGLR}) would reduce $d$ if $d<2-\sqrt{2}$ and amplify $d$ if $d>2-\sqrt{2}$. 
Generalizing to the entire graph, we see that RGLR pushes individual node pairs to the two minima and thus promotes PWS of $\mathbf{N}$ overall.

For comparison, we also analyze commonly used graph Laplacian regularizer (GLR) for surface normals. GLR for surface normals can be expressed as
\vspace{-3pt}
\begin{equation}
|\vert\mathbf{N}|\vert_{\text{GLR}}=\mathbf{n}_x^\top\mathbf{L}\mathbf{n}_x+\mathbf{n}_y^\top\mathbf{L}\mathbf{n}_y+\mathbf{n}_z^\top\mathbf{L}\mathbf{n}_z=\sum_{i,j\in\mathcal{E}}w_{i,j}|\vert\mathbf{n}_{i}-\mathbf{n}_{j}|\vert_2^2.
\label{eq:main_GLR}
\end{equation}
GLR initializes $w_{i,j}$ using (\ref{eq:edge_weight}) and keeps it fixed. 
Hence, $\mathbf{L}$ is also fixed to its initial matrix. 
With $d=\norm{\mathbf{n}_i-\mathbf{n}_j}_2^2$ and fixed $w_{i,j}$, GLR for pair $(i,j)$ is $w_{i,j}\norm{\mathbf{n}_i-\mathbf{n}_j}_2^2=w_{i,j}d$, which is a linear function of $d$ with slope $w_{i,j}$. The curve of this GLR has only one minimum at $d=0$ as shown in Fig.\;\ref{fig:normalvec}. 
Hence, considering only pair $(i,j)$, minimizing (\ref{eq:main_GLR}) only pushes $d$ towards $0$. 
Thus RGLR more aggressively promotes PWS of the target signal than GLR.
 
\vspace{-10pt}
\subsection{Problem Formulation}
\label{sec:problem}

Using our proposed RGLR (\ref{eq:main_RGLR}) as a signal prior, we formulate our point cloud denoising problem as an optimization with a general $l_p$-norm fidelity term, $\norm{\mathbf{q}-\mathbf{p}}_p^p$. Here, $\mathbf{p}$ is the optimization variable, and $\mathbf{n}_{i}$'s are functions of $\mathbf{p}$. 
Unfortunately, using state-of-the-art surface normal estimation methods~\cite{huang2001,kanatani2005,ouyang2005,gouraud1971,jin2005}, each $\mathbf{n}_{i}$ is a nonlinear function of $\mathbf{p}_{i}$ and its neighbors. 
Hence, it is difficult to formulate a tractable optimization problem using RGLR.

To overcome this difficulty, we first \textit{partition} the set of 3D points into two classes (say red and blue). 
To compute surface normal of a red point, we use only coordinates of the red point and its neighboring blue points, resulting in a linear relationship between surface normals and 3D coordinates of red points. 
Towards this goal, we compute a bipartite graph approximation of the original graph $\mathcal{G}$ as follows. 

\subsubsection{Bipartite Graph Approximation}
\label{sec:bipart}
A bipartite graph $\mathcal{B}~=~(\mathcal{V}_1,\mathcal{V}_2,\mathcal{E}')$ is a graph whose nodes are divided into two disjoint sets $\mathcal{V}_1$ and $\mathcal{V}_2$ (red and blue nodes respectively), such that each edge connects a node in $\mathcal{V}_1$ to one in $\mathcal{V}_2$. 
Given an original graph $\mathcal{G}~=~(\mathcal{V},\mathcal{E})$, our goal is to find a bipartite graph $\mathcal{B}$ that is ``closest" to $\mathcal{G}$ under a chosen metric.

First, we assume that the generative model for graph signals $\mathbf{f}$ on a given graph $\mathcal{G}$ is a \textit{Gaussian Markov random field} (GMRF) \cite{rue2005}. 
Specifically, $\mathbf{f}\sim\mathcal{N}(\mathbf{\mu},\mathbf{\Sigma})$, where $\mathbf{\mu}$ is the mean vector, and $\mathbf{\Sigma}$ is the covariance matrix specified by the graph Laplacian matrix $\mathbf{L}$ of $\mathcal{G}$. Therefore, $\Sigma^{-1}~=~\mathbf{L}+\delta\mathbf{I}$, where $1/\delta$ is the variance of the DC component for $\mathbf{f}$. 
For simplicity, we assume $\mathbf{\mu}=\mathbf{0}$.

We now find a graph $\mathcal{B}$ whose probability distribution $\mathcal{N}_B(\mathbf{0},\mathbf{\Sigma}_{\mathcal{B}})$ under the same GMRF model is closest to $\mathcal{N}(\mathbf{0},\mathbf{\Sigma})$ in terms of  \textit{Kullback-Leibler Divergence} (KLD) \cite{duchi2007}:
\begin{equation}
D_{KL}(\mathcal{N}||\mathcal{N}_{\mathcal{B}})=\frac{1}{2}(\text{tr}(\mathbf{\Sigma}_{\mathcal{B}}^{-1}\mathbf{\Sigma})+\text{ln}|\mathbf{\Sigma}_{\mathcal{B}}\mathbf{\Sigma}^{-1}|-N),
\label{eq:KLD}
\end{equation}
where $\Sigma_{\mathcal{B}}^{-1}~=~\mathbf{L}_{\mathcal{B}}+\delta\mathbf{I}$ is the precision matrix of the GMRF specified by $\mathcal{B}$, and $\mathbf{L}_{\mathcal{B}}$ is the graph Laplacian matrix of $\mathcal{B}$. To minimize (\ref{eq:KLD}), we use an iterative greedy algorithm similar to the one proposed in \cite{zeng2017} as follows.

Given a non-bipartite graph $\mathcal{G}$, we build a bipartite graph $\mathcal{B}$ by iteratively adding one node at a time to one of two disjoint sets ($\mathcal{V}_{1}$ and $\mathcal{V}_{2}$) and removing the edges within each set. 
First, we initialize one node set $\mathcal{V}_1$ with one randomly chosen node, and $\mathcal{V}_{2}$ is empty. 
Then, we use \textit{breadth-first search} (BFS) \cite{ba3na2002} to explore nodes within one hop in $\mathcal{G}$. 
To determine to which set between $\mathcal{V}_1$ and $\mathcal{V}_2$ the next node should be assigned, we calculate KLD $D_{KL}^{i}$, where $i \in \{1,2\}$, assuming the node is assigned to $\mathcal{V}_1$ or $\mathcal{V}_2$ respectively. 
The node is assigned to $\mathcal{V}_{1}$ if $D_{KL}^{2}>D_{KL}^{1}$, and to $\mathcal{V}_2$ otherwise. 
If $D_{KL}^{2}=D_{KL}^{1}$, then the node is alternately allocated to $\mathcal{V}_1$ and $\mathcal{V}_2$.

\begin{figure}[t]
\centering
\subfloat[original graph]{
\includegraphics[width=0.23\textwidth]{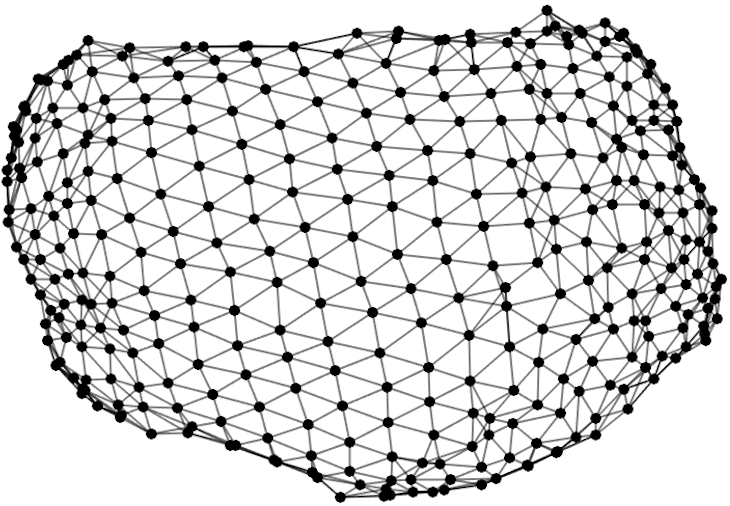}
\label{fig:pcd-day}}
\subfloat[bipartite graph]{
\includegraphics[width=0.23\textwidth]{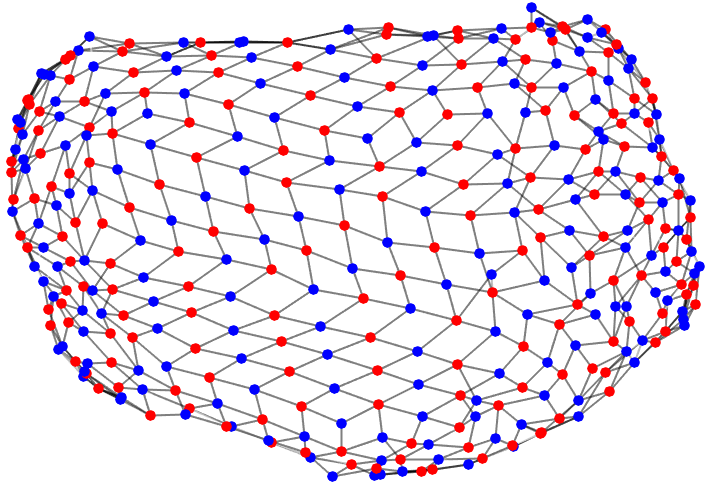}
\label{fig:pcd-on}}
\vspace{-5pt}
\caption[]{An example for bipartite graph approximation.} 
\label{fig:bipartite_graph}
\vspace{-10pt}
\end{figure}

An example of a bipartite graph approximation is shown in Fig.\;\ref{fig:bipartite_graph}. 
Here, we construct the original $k$-NN graph ($k=6$) using a small portion of the Bunny point cloud model in~\cite{levoy2005}.

\begin{figure}[t]
 \centering
 \includegraphics[width=1.1in]{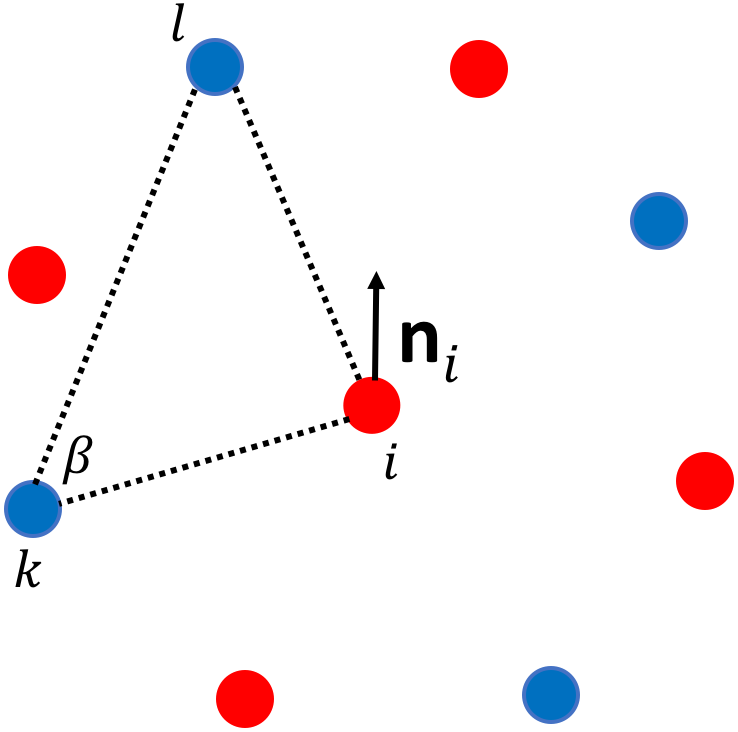} 
 \vspace{-10pt}
 \caption{Illustration of the normal vector estimation at a red node.}
 \label{fig:normalvec}
\vspace{-15pt}						
  \end{figure} 
  
\subsubsection{Normal Vector Estimation}
\label{sec:normalvector}

We describe how we define $\mathbf{n}_{i}$ for a red node $i$ of the approximated bipartite graph. For each red node $i$, we choose two connected blue nodes $k$ and $l$ (with a criterion to be discussed) in order to define a 2D plane supported by nodes $i$, $k$ and $l$. 
$\mathbf{n}_i$ is a unit-norm vector perpendicular to the plane.
See Fig.\;\ref{fig:normalvec} for an illustration. 
The corresponding 3D coordinates of nodes $i$, $k$, and $l$ are denoted by $\mathbf{p}_i~=~\begin{bmatrix} x_i & y_i & z_i \end{bmatrix}^\top$,  $\mathbf{p}^i_k~=~\begin{bmatrix} x_k & y_k & z_k \end{bmatrix}^\top$, and $\mathbf{p}^i_l~=~\begin{bmatrix} x_l & y_l & z_l \end{bmatrix}^T$ respectively. 

To select blue nodes $k$ and $l$ for a given red node $i$, we use the following two criteria: i) $\norm{\mathbf{p}_i-\mathbf{p}_k^{i}}_2$ is larger than a pre-defined threshold; and ii) the angle $\beta_i$ between $[\mathbf{p}_i-\mathbf{p}_k^{i}]$ and $[\mathbf{p}_k^{i}-\mathbf{p}_l^{i}]$ is as close to $90$ degrees as possible. 
To be discussed soon, these two criteria lead to numerical stability for our solution. Mathematically, we define $\mathbf{n}_i$ as the unit-norm vector perpendicular to that plane:
\vspace{-2pt}
\begin{equation}
\mathbf{n}_i=\frac{[\mathbf{p}_i-\mathbf{p}^i_k]\times [\mathbf{p}^i_k-\mathbf{p}^i_l]}{|\vert[\mathbf{p}_i-\mathbf{p}^i_k]\times [\mathbf{p}^i_k-\mathbf{p}^i_l]|\vert_2},
\label{eq:cross_prod}
\vspace{-0pt}
\end{equation}
\vspace{-0pt}\noindent
where `$\times$' represents the vector cross product operator. 
We rewrite the cross product in (\ref{eq:cross_prod}) as follows: 
\begin{equation}
\begin{split}
[\mathbf{p}_i-\mathbf{p}^i_k]\times [\mathbf{p}^i_k-\mathbf{p}^i_l]=\underbrace{\begin{bmatrix}0 & z_k-z_l & y_l-y_k \\
                         z_l-z_k & 0 &x_k-x_l \\
                         y_k-y_l & x_l-x_k & 0\end{bmatrix}}_{\mathbf{C}_i}\begin{bmatrix}x_i\\y_i\\z_i \end{bmatrix}\\+
                         \underbrace{\begin{bmatrix}-y_k(z_k-z_l)-z_k(y_l-y_k)\\-x_k(z_k-z_l)-z_k(x_k-x_l)\\-x_k(y_k-y_l)-y_k(x_l-x_k)\end{bmatrix}}_{\mathbf{d}_i}\\
 \end{split}
 \label{eq:linear}
 \end{equation}
 \vspace{-0pt}
Thus, normal vector $\mathbf{n}_{i}$ can be re-written as:
 \begin{equation}
 \mathbf{n}_i=\frac{\mathbf{C}_{i}\mathbf{p}_{i}+\mathbf{d}_{i}}{|\vert\mathbf{C}_{i}\mathbf{p}_{i}+\mathbf{d}_{i}|\vert_2}.
 \label{eq:norm_vec}
 \end{equation}
 
The orientations of the normal vectors for a local neighborhood of red nodes computed using (\ref{eq:norm_vec}) are not necessarily \textit{consistent}, \textit{i.e.} one normal vector might point inward from the 3D object surface while a neighoboring normal might point outward. 
To achieve consistent orientations of the surface normals, we fix the orientation of one normal vector and propagate this information to neighboring red nodes as done in ~\cite{dinesh2018fast}. 
Consistently oriented normal vectors can be written as:
\vspace{-2pt}
 \begin{equation}
 \mathbf{n}_i=\left(\frac{\mathbf{C}_{i}\mathbf{p}_{i}+\mathbf{d}_{i}}{|\vert\mathbf{C}_{i}\mathbf{p}_{i}+\mathbf{d}_{i}|\vert_2}\right)\alpha,
 \label{eq:norm_vec_ori}
 \end{equation}
where $\alpha$ is 1 or $-1$ depending on the normal orientation.

For simplicity we assume that $\alpha$ and $|\vert\mathbf{C}_{i}\mathbf{p}_{i}+\mathbf{d}_{i}|\vert_2$ remain constant when $\mathbf{p}_{i}$ is being optimized. 
Hence, surface normal $\mathbf{n}_i$ can be written as a linear function of 3D coordinates $\mathbf{p}_i$ of the point $i$. 
Specifically,
\begin{equation}
\mathbf{n}_{i}=\mathbf{A}_{i}\mathbf{p}_{i}+\mathbf{b}_{i},
\label{eq:normal_linear}
\end{equation}
where $\mathbf{A}_{i}~=~\mathbf{C}_{i}\alpha^{in}/|\vert\mathbf{C}_{i}\mathbf{p}_{i}^{in}+\mathbf{d}_{i}|\vert_2$ and $\mathbf{b}_{i}~=~\mathbf{d}_{i}\alpha^{in}/|\vert\mathbf{C}_{i}\mathbf{p}_{i}^{in}+\mathbf{d}_{i}|\vert_2$. Here $\mathbf{p}_{i}^{in}$ is the initial vector of $\mathbf{p}_{i}$ and $\alpha^{in}$ is the value of $\alpha$ computed at $\mathbf{p}_{i}=\mathbf{p}_{i}^{in}$.

\subsubsection{Optimization Cost}

After computing surface normals for each red node based on blue nodes, we construct a \textit{new} $k$-NN graph $\mathcal{G}_{\mathcal{R}}=(\mathcal{V}_{\mathcal{R}},\mathcal{E}_{\mathcal{R}})$ for red nodes, where $\mathcal{V}_{\mathcal{R}}$ is the set of red nodes and $\mathbf{\mathcal{E}_{\mathcal{R}}}$ is the set of edges connecting neighboring red nodes. 
Given $\mathcal{G}_{\mathcal{R}}$, we define an $\ell_p$-$\ell_2$-norm minimization problem for denoising:
\begin{equation}
\min_{\mathbf{p}}|\vert\mathbf{q-p}\vert|_{p}^{p} + \gamma|\vert\mathbf{N}|\vert_{\text{RGLR}}
\label{eq:main_opt}
\end{equation}
where $\gamma$ is a weight parameter that trades off the fidelity term $|\vert\mathbf{q-p}\vert|_{p}^{p}$ with prior $\norm{\mathbf{N}}_{\text{RGLR}}$ (\ref{eq:main_RGLR}). 
Small but non-sparse noise (like Gaussian) is conventionally modeled using an $\ell_2$-norm fidelity term~\cite{chambolle2009, chan1999, daubechies2006}. Further, large but sparse noise (like Laplacian) can be modeled using a convex $\ell_1$-norm fidelity term instead~\cite{nikolova2004, alliney1997}. Depending on the point cloud acquisition method used, one of the two noise models can be appropriately chosen. We focus on solving the denoising problem in~(\ref{eq:main_opt}) explicitly for these two noise models.

According to (\ref{eq:normal_linear}), we can write $\mathbf{n}_x$, $\mathbf{n}_y$, and $\mathbf{n}_z$ used in $\norm{\mathbf{N}}_{\text{RGLR}}$ for $\mathcal{G}_{\mathcal{R}}$ as follows:
\vspace{-7pt}
\begin{equation}
\begin{split}
\mathbf{n}_x=\underbrace{\text{diag}\{\mathbf{A}_1(1,:),\hdots,\mathbf{A}_N(1,:)\}}_{\mathbf{A}_x}\mathbf{p}+\underbrace{\begin{bmatrix}
b_{11} & \hdots & b_{N1}
\end{bmatrix}^\top}_{\mathbf{b}_x},\\
\mathbf{n}_y=\underbrace{\text{diag}\{\mathbf{A}_1(2,:),\hdots,\mathbf{A}_N(2,:)\}}_{\mathbf{A}_y}\mathbf{p}+\underbrace{\begin{bmatrix}
b_{12} & \hdots & b_{N2}
\end{bmatrix}^\top}_{\mathbf{b}_y},\\
\mathbf{n}_z=\underbrace{\text{diag}\{\mathbf{A}_1(3,:),\hdots,\mathbf{A}_N(3,:)\}}_{\mathbf{A}_z}\mathbf{p}+\underbrace{\begin{bmatrix}
b_{13} & \hdots & b_{N3}
\end{bmatrix}^\top}_{\mathbf{b}_z},
\end{split}
\label{eq:nx_ny_nz}
\end{equation}
where $\mathbf{A}_{i}(j,:)$ denotes the $j$-th row of $\mathbf{A}_i$ and $b_{ij}$ denotes the $j$-th entry of $\mathbf{b}_i$. 
Given (\ref{eq:nx_ny_nz}), we rewrite $\norm{\mathbf{N}}_{\text{RGLR}}$ as
\begin{equation}
\begin{bmatrix} \mathbf{A}_x\mathbf{p}+\mathbf{b}_x\\
\mathbf{A}_y\mathbf{p}+\mathbf{b}_y\\
\mathbf{A}_z\mathbf{p}+\mathbf{b}_z
\end{bmatrix}^\top\begin{bmatrix}\mathbf{L(p)} &\mathbf{0} & \mathbf{0}\\
\mathbf{0}& \mathbf{L(p)} &\mathbf{0}\\
\mathbf{0} & \mathbf{0} & \mathbf{L(p)}
\end{bmatrix} \begin{bmatrix} \mathbf{A}_x\mathbf{p}+\mathbf{b}_x\\
\mathbf{A}_y\mathbf{p}+\mathbf{b}_y\\
\mathbf{A}_z\mathbf{p}+\mathbf{b}_z
\end{bmatrix}.
\label{eq:l2_l2_optimization}
\end{equation}
Since $\mathbf{n}$ is a function of $\mathbf{p}$, we write $\mathbf{L}(\mathbf{N})$ as $\mathbf{L}(\mathbf{p})$ instead in the sequel. 

In our optimization framework, we first solve (\ref{eq:main_opt}) for red nodes while keeping the positions of blue nodes fixed. 
Then, using the newly solved red nodes' positions, we compute the surface normal for each blue node and construct another $k$-NN graph $\mathcal{G}_{\mathcal{B}}(\mathbf{V}_{\mathcal{B}},\mathcal{E}_{\mathcal{B}})$ for blue nodes. Then (\ref{eq:main_opt}) is solved for blue nodes while keeping the positions of red nodes fixed. 
The two node sets are alternately optimized until convergence. 
We next discuss our optimization algorithm to solve (\ref{eq:main_opt}) for a given set of nodes (say red nodes in graph $\mathcal{G}_{\mathcal{R}}$) at one iteration.

%% file: optimization.tex
Since $\mathbf{L}(\mathbf{p})$ is a function of $\mathbf{p}$, optimization~(\ref{eq:main_opt}) is non-convex, which is challenging to solve. 
Thus, we use an iterative method to optimize $\mathbf{p}$. 
At the $k$-th iteration, we fix $\mathbf{L}~=~\mathbf{L}\left(\mathbf{p}^{(k-1)}\right)$, then compute $\mathbf{p}^{(k)}$ by minimizing (\ref{eq:main_opt}). 
Here, we initialize $\mathbf{p}^{(0)}~=~\mathbf{q}$. With fixed $\mathbf{L}$, (\ref{eq:main_opt}) becomes a convex optimization problem for $p \in \{1, 2\}$:
\begin{equation}
\min_{\hat{\mathbf{p}}}\norm{\mathbf{q}-\hat{\mathbf{p}}}_p^p+\gamma\underbrace{\begin{bmatrix} \mathbf{L}.(\mathbf{A}_x\hat{\mathbf{p}}+\mathbf{b}_x)\\
\mathbf{L}.(\mathbf{A}_y\hat{\mathbf{p}}+\mathbf{b}_y)\\
\mathbf{L}.(\mathbf{A}_z\hat{\mathbf{p}}+\mathbf{b}_z)
\end{bmatrix}^{\top}\begin{bmatrix} \mathbf{A}_x\hat{\mathbf{p}}+\mathbf{b}_x\\
\mathbf{A}_y\hat{\mathbf{p}}+\mathbf{b}_y\\
\mathbf{A}_z\hat{\mathbf{p}}+\mathbf{b}_z
\end{bmatrix}}_{f(\hat{\mathbf{p}})}.
\label{eq:lp_l2_optimization}
\end{equation}
\vspace{-20pt}
\subsection{Solving (\ref{eq:lp_l2_optimization}) with $\ell_2$-norm fidelity term}
\label{sec:sol for l1}

When $p~=~2$, (\ref{eq:lp_l2_optimization}) becomes a minimization of a quadratic cost of $\hat{\mathbf{p}\in\mathbb{R}^{3N_{\mathcal{R}}}}$ ($N_{\mathcal{R}}$ is the number of nodes in the graph $\mathcal{G}_{\mathcal{R}}$):
\vspace{-5pt}
\begin{equation}
\min_{\hat{\mathbf{p}}}\norm{\mathbf{q}-\hat{\mathbf{p}}}_2^2+\gamma f(\hat{\mathbf{p}}).
\label{eq:l2_l2_optimization}
\end{equation}
To optimize $\hat{\mathbf{p}}$, we first take the derivative of (\ref{eq:l2_l2_optimization}) with respect to $\hat{\mathbf{p}}$, set it to $\mathbf{0}$ and solve for the closed form solution $\mathbf{p}^k$:
\begin{equation}
\begin{bmatrix}
\mathbf{I}+\gamma\underbrace{\begin{bmatrix} \mathbf{LA}_x\\
\mathbf{LA}_y\\
\mathbf{LA}_z
\end{bmatrix}^{\top}\begin{bmatrix} \mathbf{A}_x\\
\mathbf{A}_y\\
\mathbf{A}_z
\end{bmatrix}}_{\tilde{\mathbf{L}}}
\end{bmatrix}\mathbf{p}^{(k)}=\mathbf{q}-\gamma\underbrace{\begin{bmatrix} \mathbf{LA}_x\\
\mathbf{LA}_y\\
\mathbf{LA}_z
\end{bmatrix}^{\top}\begin{bmatrix} \mathbf{b}_x\\
\mathbf{b}_y\\
\mathbf{b}_z
\end{bmatrix}}_{\bar{\mathbf{L}}},
\label{eq:sol_qud}
\end{equation}
where $\mathbf{I}$ is an identity matrix. Since $\mathbf{L}$ is a positive semi-definite (PSD) matrix, $\tilde{\mathbf{L}}$ is also PSD. 
Hence~$\mathbf{I}+\gamma\tilde{\mathbf{L}}$ is positive definite (PD) for~$\gamma>0$ and thus invertible. 
We demonstrate soon that solving the system of linear equations in~(\ref{eq:sol_qud}) is numerically stable by showing $\mathbf{I}+\gamma\tilde{\mathbf{L}}$ has a bounded \textit{condition number}\footnote{The condition number of a given square symmetric matrix is defined as the ratio of its maximum and minimum eigenvalues.}. 
Further, $\mathbf{I}+\gamma\tilde{\mathbf{L}}$ is a sparse matrix. 
Hence we can efficiently solve~(\ref{eq:sol_qud}) using \textit{conjugate gradient} (CG)~\cite{shewchuk1994} without full matrix inversion. 

After obtaining $\mathbf{p}^{(k)}$ from (\ref{eq:sol_qud}), $\mathbf{L}$ is updated as $\mathbf{L}~=~\mathbf{L}(\mathbf{p}^{(k)})$ and then $\mathbf{p}^{(k+1)}$ is computed using (\ref{eq:sol_qud}) again. Repeating this procedure, $\mathbf{p}$ is iteratively optimized until convergence. 


\subsubsection{Numerical Stability}
\label{se:stability}
We demonstrate the numerical stability of the system of linear equations in~(\ref{eq:sol_qud}) based on the condition number of $\mathbf{I}+\gamma\tilde{\mathbf{L}}$ via the following eigen-analysis. 
Denote the maximum eigenvalues of $\mathbf{A}_x^\top\mathbf{L}\mathbf{A}_x$, $\mathbf{A}_y^\top\mathbf{L}\mathbf{A}_y$ and $\mathbf{A}_z^\top\mathbf{L}\mathbf{A}_z$ by $\mu_x^{\max}$, $\mu_y^{\max}$ and $\mu_z^{\max}$, respectively. 
According to \textit{Weyl's inequality in matrix theory}\footnote{https://en.wikipedia.org/wiki/Weyl\%27s\_inequality}, we know that $\mu^{\max}~\leq~ \mu_x^{\max}+\mu_y^{\max}+\mu_z^{\max}$, where $\mu^{\max}$ is the maximum eigenvalue of $\tilde{\mathbf{L}}~=~\mathbf{A}_x^\top\mathbf{L}\mathbf{A}_x+\mathbf{A}_y^\top\mathbf{L}\mathbf{A}_y+\mathbf{A}_z^\top\mathbf{L}\mathbf{A}_z$. 
Moreover, since $\tilde{\mathbf{L}}$ is PSD, its minimum eigenvalue is larger or equal to $0$. We see easily that the maximum eigenvalue of $\mathbf{I}+\gamma\tilde{\mathbf{L}}$ is upper bounded by $1+\gamma\mu^{\max}$ and its minimum eigenvalue is lower bounded by $1$. 
Thus we can upper-bound condition number $C$ for $\mathbf{I}+\gamma\tilde{\mathbf{L}}$ as follows:
\vspace{-2pt}
\begin{equation}
C\leq1+\gamma(\mu_x^{\max}+\mu_y^{\max}+\mu_z^{\max}).
\label{eq:cond}
\vspace{-1pt}
\end{equation}

We derive an upper bound for $\mu_x^{\max}$. 
Denote by $\mathbf{v}$ the unit-norm eigenvector corresponding to $\mu_x^{\max}$. 
By the definition of eigenvalues and eigenvectors, $\mathbf{v}^\top\mathbf{A}_x^\top\mathbf{L}\mathbf{A}_x\mathbf{v}~=~\mu_x^{\max}$. 
Denote the maximum eigenvalue of $\mathbf{L}$ by $\lambda^{\max}$. 
Following the \textit{min-max theorem}\footnote{https://en.wikipedia.org/wiki/Min-max\_theorem}, we know that:
\begin{equation}
\frac{\mathbf{u}^\top\mathbf{L}\mathbf{u}}{\norm{\mathbf{u}}_2^2}\leq \lambda^{\max}, \hspace{8pt} \forall \mathbf{u}\neq\mathbf{0}.
\label{eq:minmax}
\end{equation}
If we select $\mathbf{u}~=~\mathbf{A}_x\mathbf{v}$ and combine~(\ref{eq:minmax}) with the expression $\mathbf{v}^\top\mathbf{A}_x^\top\mathbf{L}\mathbf{A}_x\mathbf{v}~=~\mu_x^{\max}$, we can write:
\begin{equation}
\begin{split}
\mu_x^{\max}&\leq\norm{\mathbf{A}_x\mathbf{v}}_2^2\lambda^{\max}\leq \norm{\mathbf{A}_x}_2^2\norm{\mathbf{v}}_2^2\lambda^{\max}\\ &=\norm{\mathbf{A}_x}_2^2\lambda^{\max},
\end{split}
\end{equation}
where $\norm{\mathbf{v}}_2^2~=~1$. 

Moreover, using \textit{Gershgorin circle theorem}~\cite{brakken2007}, we can prove that $\lambda^{\max}\leq 2\rho^{\max}$, where $\rho^{\max}$ is the maximum degree of a node in the $k$-NN graph $\mathcal{G}_{\mathcal{R}}$ (our proof is given in section 1.A in the supplement). 
Hence $\mu_x^{\max}~\leq~ 2\rho^{\max}\norm{\mathbf{A}_x}_2^2$, and we can similarly derive $\mu_y^{\max}~\leq~ 2\rho^{\max}\norm{\mathbf{A}_y}_2^2$ and $\mu_z^{\max}~\leq~ 2\rho^{\max}\norm{\mathbf{A}_z}_2^2$. We now rewrite (\ref{eq:cond}) as follows: 
\begin{equation}
C\leq 1+2\gamma\rho^{\max}\left(\norm{\mathbf{A}_x}_2^2+\norm{\mathbf{A}_y}_2^2+\norm{\mathbf{A}_z}_2^2\right).
\label{eq:conditinnumber}
\end{equation}

Note that $\norm{\mathbf{A}_x}_2^2$ is equivalent\footnote{https://en.wikipedia.org/wiki/Matrix\_norm} to the largest eigenvalue of $\mathbf{A}_x\mathbf{A}_x^{\top}$. 
By the definition of $\mathbf{A}_x$ in~(\ref{eq:nx_ny_nz}), $\mathbf{A}_x\mathbf{A}_x^{\top}$ is  a diagonal matrix with its $(i,i)$ entry equal\footnote{The value of the entry $(i,i)$ of $\mathbf{A}_x\mathbf{A}_x^{\top}$ is obtained by following~(\ref{eq:cross_prod}),~(\ref{eq:linear}),~and~(\ref{eq:norm_vec_ori}). By definition $\norm{[\mathbf{p}_i-\mathbf{p}_k^{i}]\times[\mathbf{p}_k^{i}-\mathbf{p}_l^{i}]}~=~\norm{\mathbf{p}_k^{i}-\mathbf{p}_l^{i}}_2\norm{\mathbf{p}_i-\mathbf{p}_k^{i}}_2\sin\beta_{i}$.} to 
\begin{equation}
\frac{(z_k-z_l)^2+(y_l-y_k)^2}{\norm{\mathbf{p}_k^{i}-\mathbf{p}_l^{i}}_2^2\norm{\mathbf{p}_i-\mathbf{p}_k^{i}}_2^2\sin^2\beta_{i}}\leq\frac{1}{\norm{\mathbf{p}_i-\mathbf{p}_k^{i}}_2^2\sin^2\beta_i}.
\label{eq:diginaleliments}
\end{equation}
Hence we can write an upper bound for $\norm{\mathbf{A}_x}_2^2$ (\textit{i.e.} the maximum eigenvalue of $\mathbf{A}_x\mathbf{A}_x^{\top}$) as follows:
\begin{equation}
\norm{\mathbf{A}_x}_2^2\leq\frac{1}{\min_i \norm{\mathbf{p}_i-\mathbf{p}_k^{i}}_2^2\sin^2\beta_i}
\label{eq:openedUper}
\end{equation}
We see that $\norm{\mathbf{A}_y}_2^2$ and $\norm{\mathbf{A}_z}_2^2$ also have the same upper bound. 
From (\ref{eq:conditinnumber}), we can write an upper bound for $C$ as follows:
\vspace{-5pt}
\begin{equation}
C\leq 1+\frac{6\gamma\rho^{\max}}{\min_i \norm{\mathbf{p}_i-\mathbf{p}_k^{i}}_2^2\sin^2\beta_i}.
\label{eq:finalC}
\end{equation}

For sufficiently small $\rho^{\max}$ and $\gamma$, and for sufficiently large $\min_i \norm{\mathbf{p}_i-\mathbf{p}_k^{i}}_2^2\sin^2\beta_i$, $C$ is sufficiently small, and thus the system equations in~(\ref{eq:sol_qud}) has a numerically stable solution. 
Typically, in our experiments $\rho^{\max}<10$ and $\gamma<0.8$. 
Moreover, we select neighboring blue nodes $k$ and $l$ for a given red node $i$ so that $\norm{\mathbf{p}_i-\mathbf{p}_k^{i}}_2$ is sufficiently larger and $\beta_i$ is close to $90$ degrees, as discussed in Section~\ref{sec:normalvector}. 
Therefore, in practice, $C$ has a sufficiently small value so that the system equations in~(\ref{eq:sol_qud}) has a stable solution.


\subsubsection{Graph spectral interpretation to solve (\ref{eq:sol_qud})}
According to (\ref{eq:sol_qud}), we can obtain $\mathbf{p}^{(k)}$ as follows:
\begin{equation}
\mathbf{p}^{(k)}=(\mathbf{I}+\gamma\tilde{\mathbf{L}})^{-1}\tilde{\mathbf{q}},
\label{eq:graph_filter}
\end{equation}
where $\tilde{\mathbf{q}}~=~\mathbf{q}-\gamma\bar{\mathbf{L}}$. Denote by $0\leq \mu_1\leq \hdots \leq\mu_N$ the eigenvalues of $\tilde{\mathbf{L}}$ and $\mathbf{\phi}_1, \hdots, \mathbf{\phi}_N$ the corresponding eigenvectors. 
The solution to (\ref{eq:graph_filter}) can thus be written as:
\begin{equation}
\mathbf{p}^{(k)}=\mathbf{\Phi\Sigma}^{-1}\mathbf{\Phi}^\top\tilde{\mathbf{q}},
\label{eq:spectral}
\end{equation}
where $\mathbf{I}+\gamma\tilde{\mathbf{L}}~=~\mathbf{\Phi\Sigma}\mathbf{\Phi}^{\top}$, which is the orthogonal decomposition of $\mathbf{I}+\gamma\tilde{\mathbf{L}}$ with $\mathbf{\Phi}~=~\begin{bmatrix}
\phi_1 & \hdots & \phi_N
\end{bmatrix}$ and $\mathbf{\Sigma}~=~\text{diag}(1+\gamma\mu_1, \hdots, 1+\gamma\mu_N)$. In GSP, eigenvalues and eigenvectors of the Laplacian matrix are commonly interpreted as graph frequencies and frequency basis~\cite{shuman2013}. 
Analogously, here we interpret $\mathbf{\Phi}^\top$ as a \textit{graph Fourier transform} (GFT) operator that maps a graph signal $\mathbf{f}$ to its GFT coefficients $\mathbf{\Phi}^{\top}\mathbf{f}$. Therefore, observing that $\mathbf{\Sigma}^{-1}~=~\text{diag}(1/(1+\gamma\mu_{1}), \hdots, 1/(1+\gamma\mu_{N}))$, we can interpret the solution to $\mathbf{p}^{(k)}$ in~(\ref{eq:spectral}) as follows. 
The graph signal $\tilde{\mathbf{q}}$ is transformed to the frequency domain via $\mathbf{\Phi}^{\top}$, attenuated GFT coefficients according to $\mathbf{\Sigma}^{-1}$ and transformed back to the nodal domain via $\mathbf{\Phi}$. 
This is called \textit{graph spectral filtering}~(GSF)~\cite{shuman2013}. We can thus interpret solving~(\ref{eq:graph_filter}) as a \textit{low-pass} filtering (since the attenuation is stronger for larger eigenvalues) of graph signal $\tilde{\mathbf{q}}$ using a polynomial graph filter $g(\tilde{\mathbf{L}})~=~(\mathbf{I}+\gamma\tilde{\mathbf{L}})^{-1}$.

For graph spectral filters, we can solve the graph filtering problem in (\ref{eq:graph_filter}) using an accelerated filter on graphs via the \textit{Lanczos method}\cite{susnjara2015} as an alternative to CG. 
The Lanczos method computes an orthonormal basis $\mathbf{V}_{M}~=~\begin{bmatrix}
\mathbf{v}_1 \hdots \mathbf{v}_M
\end{bmatrix}$ of the Krylov subspace $\mathbf{K}_{M}(\tilde{\mathbf{L}},\tilde{\mathbf{q}})~=~\text{span}\left\{\tilde{\mathbf{q}}, \tilde{\mathbf{L}}\tilde{\mathbf{q}}, \hdots, \tilde{\mathbf{L}}^{M-1}\tilde{\mathbf{q}} \right\}$ and the corresponding symmetric tridiagonal matrix $\mathbf{H}_M\in \mathbb{R}^{M\times M}$ satisfying $\mathbf{V}_{M}^{\text{H}}\tilde{\mathbf{L}}\mathbf{V}_{M}\approx\mathbf{H}_M$, where the exact non-zero entities of $\mathbf{H}_M$ can be computed using Algorithm 1 in~\cite{susnjara2015}. 
The approximation of $\mathbf{p}^{(k)}$ with an order-$M$ Lanczos method, where $M \ll N$, is
\vspace{-7pt}
\begin{equation}
\mathbf{p}^{(k)}=g(\tilde{\mathbf{L}})\tilde{\mathbf{q}}\approx\norm{\tilde{\mathbf{q}}}_2\mathbf{V}_Mg({\mathbf{H}_M})\mathbf{e}_1,
\end{equation}
where $\mathbf{e}_1\in \mathbb{R}^{M}$ is the first \textit{canonical vector} (all elements are zeros except the first element is one), and $g({\mathbf{H}_M})~=~(\mathbf{I}+\gamma\mathbf{H}_{M})^{-1}$. 
Given $M \ll N$ and since $\mathbf{I}+\gamma\mathbf{H}_{M}$ (denoted as $\tilde{g}(\mathbf{H}_M)$) is a symmetric tridiagonal matrix, orthogonal decomposition of $\tilde{g}(\mathbf{H}_M)$ can be computed using a \textit{fast divide-and-conquer algorithm}~\cite{coakley2013} with computational complexity of $\mathcal{O}(M\ln M)$. Then we can directly obtain $g({\mathbf{H}_M})$ using the orthogonal decomposition of $\tilde{g}({\mathbf{H}_M})$.      
\vspace{-10pt}
\subsection{Solving (\ref{eq:lp_l2_optimization}) with $\ell_1$-norm fidelity term}
\label{sec:l2fid}

When $p~=~1$, (\ref{eq:main_opt}) becomes a $\ell_1$-$\ell_2$ minimization problem as follows:
\vspace{-7pt}
\begin{equation}
\min_{\hat{\mathbf{p}}}\norm{\mathbf{q}-\hat{\mathbf{p}}}_1+\gamma{f(\hat{\mathbf{p}})},
\label{eq:l1_l2_optimization}
\end{equation}
where the second term $f(\hat{\mathbf{p}})$ is convex and differentiable, and the first term is convex but non-differentiable. 
We can thus employ \textit{accelerated proximal gradient}~(APG)~\cite{parikh2014} to solve (\ref{eq:l1_l2_optimization}). 
Differentiable $f(\hat{\mathbf{p}})$ has gradient $\nabla f$:
\begin{equation}
\nabla f(\hat{\mathbf{p}})=2\gamma(\tilde{\mathbf{L}}\hat{\mathbf{p}}+\bar{\mathbf{L}}),
\label{eq: f_grad}
\end{equation}
where $\tilde{\mathbf{L}}$ and $\bar{\mathbf{L}}$ are matrices defined in (\ref{eq:sol_qud}). 
We now define a \textit{proximal mapping} $\text{prox}_{h,t}(\hat{\mathbf{p}})$ for a convex, non-differentiable function $h()$ with step size $t$ as:
\begin{equation}
\text{prox}_{h,t}(\hat{\mathbf{p}})=\text{arg}\min_{\theta}\left\{h(\theta)+\frac{1}{2t}|\vert\theta-\hat{\mathbf{p}}\vert|_{2}^{2}\right\}
\end{equation}
It is known that if $h()$ is simply a $l_{1}$-norm in (\ref{eq:l1_l2_optimization}), then the proximal mapping is a soft thresholding function~\cite{parikh2014}:
\begin{equation}
\text{prox}_{h,t}(\hat{p}_{i})=\begin{cases}
\hat{p}_{i}-t &\text{if} \hspace{5pt} \hat{p}_{i}>q_i+t\\
q_i &\text{if} \hspace{5pt} \vert \hat{p}_{i}\vert\leq q_i+t\\
\hat{p}_{i}+t &\text{if} \hspace{5pt} \hat{p}_{i}<-q_i-t,
\end{cases}
\end{equation}
where $\hat{p}_{i}$ and $q_i$ are $i$-th entries of $\hat{\mathbf{p}}$ and $\mathbf{q}$ respectively. To adopt an APG approach, we define (with initial point $\hat{\mathbf{p}}^{(-1)}~=~\hat{\mathbf{p}}^{(0)}~=~\mathbf{p}^{(k-1)}$)
\vspace{-5pt}
\begin{equation}
\mathbf{z}^{(m-1)}=\hat{\mathbf{p}}^{(m-1)}+\frac{m-2}{m+1}\left(\hat{\mathbf{p}}^{(m-1)}-\hat{\mathbf{p}}^{(m-2)}\right),
\end{equation}
where $\mathbf{z}^{(m-1)}$ is an extrapolated point computed from the two previous solutions $\hat{\mathbf{p}}^{(m-1)}$ and $\hat{\mathbf{p}}^{(m-2)}$. We now update $\hat{\mathbf{p}}^{(m)}$ using $\mathbf{z}^{(m-1)}$ as follows:
\vspace{-5pt}
\begin{equation}
\hat{\mathbf{p}}^{(m)}=\text{prox}_{h,t}(\mathbf{z}^{(m-1)}-t\nabla f(\mathbf{z}^{(m-1)})).
\label{eq:m_update}
\end{equation}
We compute (\ref{eq:m_update}) iteratively until convergence. 

An appropriate step size $t$ for faster convergence can be selected based on \textit{Lipschitz continuity}~\cite{combettes2005} of $\nabla f$ as follows. 
We say that $\nabla f$ is Lipschitz continuous if there is a real positive constant $K$ such that, for all real $\mathbf{p}_1, \mathbf{p}_2 \in \mathbb{R}^{3N_{\mathcal{R}}}$,
\begin{equation}
\norm{\nabla f(\mathbf{p}_1)-\nabla f(\mathbf{p}_1)}_2\leq K\norm{\mathbf{p}_1-\mathbf{p}_2}_2.
\label{eq: Lip_conti}
\end{equation}
The smallest possible $K$ that satisfies (\ref{eq: Lip_conti}) is called the Lipschitiz constant $L$. 
In our specific case, according to (\ref{eq: f_grad}) we can write
\vspace{-7pt}
\begin{equation}
\begin{split}
\norm{\nabla f(\mathbf{p}_1)-\nabla f(\mathbf{p}_1)}_2 &=2\gamma\norm{\tilde{\mathbf{L}}\mathbf{p}_1-\tilde{\mathbf{L}}\mathbf{p}_1}_2\\
&\leq 2\gamma\norm{\tilde{\mathbf{L}}}_2\norm{\mathbf{p}_1-\mathbf{p}_1}_2
\end{split}
\end{equation}
for all real $\mathbf{p}_1, \mathbf{p}_2$, and hence $\nabla f$ is Lipschitz continuous with $L=2\gamma\norm{\tilde{\mathbf{L}}}_2$. Here, $\norm{\tilde{\mathbf{L}}}_2$ is the largest eigenvalue of $\tilde{\mathbf{L}}$ (since $\tilde{\mathbf{L}}$ is a symmetric matrix), and we know that it is upper-bounded by $\frac{6\rho^{max}}{\min_i \norm{\mathbf{p}_i-\mathbf{p}_k^{i}}_2^2\sin^2\beta_i}$\footnote{As stated in Section~\ref{sec:normalvector}, when selecting $k$ and $l$ blue nodes for surface normal estimation at red node $i$, we calculate $\norm{\mathbf{p}_i-\mathbf{p}_k^{i}}_2$ and $\beta_i$.} from eigen-analysis in Section~\ref{se:stability}.
As discussed in~\cite{parikh2014}, APG has convergence rate $\mathcal{O}(1/k^2)$ when a step size $t\in (0,1/L]$ is used. 
This is theoretically the fastest possible convergence rate for first-order methods.

Here, $\mathbf{p}^{(k)}$ equals to the converged $\hat{\mathbf{p}}$ obtained from~(\ref{eq:m_update}). After obtaining $\mathbf{p}^{(k)}$, $\mathbf{L}$ is updated as $\mathbf{L}~=~\mathbf{L}(\mathbf{p}^{(k)})$ and then $\mathbf{p}^{(k+1)}$ is computed based on the updated $\mathbf{L}$. Repeating this procedure, $\mathbf{p}$ is iteratively optimized until convergence in order to solve~(\ref{eq:main_opt}) with the $\ell_1$-norm fidelity term. 


%% file: estimate.tex
\subsection{Weight Parameter and Noise Variance}

The computation of an optimal weight parameter ($\gamma_{\mathrm{opt}}$) that trades off the fidelity term with the signal prior (\ref{eq:main_opt}) is in general an open problem. 
Recently, the analysis of $\gamma_{\mathrm{opt}}$ for GLR was studied in~\cite{chen2017} assuming that the ground truth signal is known, which is not practical. 

Intuitively, one can see (and we verified through extensive experiments) that $\gamma_{\mathrm{opt}}$ is proportional to the noise variance ($\sigma^2$) of an iid noise that corrupts 3D points in a given point cloud; \textit{i.e.}, the larger the noise variance, the larger the importance of the prior term.
We can thus write a simple linear relationship between $\gamma_{\mathrm{opt}}$ and $\sigma^2$ as follows:
\vspace{-5pt}
\begin{equation}
\gamma_{\mathrm{opt}} \approx \alpha \sigma^2,
\label{eq:relation_gam_sig}
\vspace{-5pt}
\end{equation}
where $\alpha$ is a pre-determined constant invariant to different point clouds corrupted by different noise levels. However, $\sigma^2$ is unknown in general. 
In the point cloud denoising literature~\cite{avron2010,mattei2017,sun2015,schoenenberger2015,dinesh2018fast}, typically $\sigma^2$ is assumed known or $\gamma_{\mathrm{opt}}$ is tuned experimentally for best performance---neither is realizable in practice when only a noisy point cloud is given as input.

In contrast, we propose a simple and efficient method to estimate $\sigma^2$ in an observed point cloud. 
Our approach is analogous to \cite{wu2015} that estimates the noise variance of natural images. Assuming an iid noise model, \cite{wu2015} proposes to: i) identify \textit{flat} patches (each patch has approximately a constant pixel value) from the given image, and ii) estimate the noise variance locally in each flat patch. 
Analogously, we first identify approximately flat 3D patches in a given noisy point cloud, and then estimate the noise variance from each patch. Before all these steps, we perform bipartite graph approximation as discussed in Section~\ref{sec:bipart}, then estimate surface normals for one partite of nodes using our method in Section~\ref{sec:normalvector}. 


To find flat patches, we first divide the point cloud into groups of 3D points with two similar features---geometric positions and estimated surface normals. 
Towards this goal, we employ a \textit{mean shift} clustering algorithm \cite{comaniciu2002} based on these two features hierarchically. 
The main reason is that mean shift is nonparametric; \textit{i.e.}, we impose no prior on the number or shapes of the clusters. 
Similar hierarchical mean shift clustering techniques have been used in \cite{zhang2018} for point cloud segmentation. 

Given a point cloud, we first perform clustering in the surface normal space. We next apply local clustering for each surface normal cluster in the geometry space again using mean shift. 
The points in the final clusters thus have similar features in both geometry space and surface normal space. 
We know that a cluster from a flat region has a larger number of similar surface normals than a cluster from a non-flat region. 
Thus, if the number of points of a given final cluster is larger than a pre-specified threshold (in our experiments, we use 25), we consider that cluster a flat patch. 
For a given flat 3D patch, we have a strong prior that the normals of all points in the patch are approximately the same, which leads to simple and fast noise variance estimation methods as discussed next.


Our noise variance estimation method has two versions that are applicable to two different kinds of noise. For Gaussian noise, we employ a \textit{mean filter} to compute first the average noise variance $\sigma_\mathbf{n}^2$ of surface normals in flat patches, then derive $\sigma^2$ from computed $\sigma_\mathbf{n}^2$. 
For Laplacian noise, we employ a \textit{median filter} to estimate $\sigma^2$ from noisy points.  
\vspace{-7pt}


\subsection{Noise Variance Estimation for Gaussian Noise}
\subsubsection{Computing Surface Normal Noise Variance $\sigma_{\mathbf{n}}$}

Given a roughly flat patch with $n_{r}$ red and $n_{b}$ blue points, we estimate the average noise variance of the surface normals for the red point set $\mathcal{S}_{r}$ in the patch as follow.
Given (\ref{eq:normal_linear}), we can write the noise corrupted surface normals $\tilde{\mathbf{n}}$ in $\mathcal{S}_{r}$, \textit{i.e.}, $\tilde{\mathbf{n}} = \left[ \tilde{\mathbf{n}}_{1}^{\top}  \ldots \tilde{\mathbf{n}}_{n_r}^{\top} \right]^{\top}$, as
\vspace{-8pt}
\begin{equation}
\tilde{\mathbf{n}}=\mathbf{A}(\mathbf{p}+\mathbf{w})+\mathbf{b}.
\label{eq:normal_vector_noise}
\end{equation}
where $\mathbf{A} ~=~ \text{diag}(\mathbf{A}_1, \hdots, \mathbf{A}_{n_r})$ and $\mathbf{b} ~=~ \left[ \mathbf{b}_{1}^{\top} \ldots \mathbf{b}_{n_r}^{\top} \right]^{\top}$. 
$\mathbf{p} ~=~ \left[ \mathbf{p}_{1}^{\top} \ldots \mathbf{p}_{n_r}^{\top} \right]^{\top}$ is the ground truth 3D coordinate vector for points in $\mathcal{S}_{r}$, and $\mathbf{w}\in \mathbb{R}^{3n_r}$ is the iid additive noise corrupting $\mathbf{p}$. 
Denote by $\mathbf{n}^o$ the surface normals for ground truth 3D points in $\mathcal{S}_{r}$,
\vspace{-10pt}
\begin{equation}
\mathbf{n}^o=\mathbf{Ap}+\mathbf{b},
\label{eq:normal_vector}
\vspace{-5pt}
\end{equation}

The average noise variance $\sigma_{\mathbf{n}}^2$---noise variance per coordinate of each 3D point---for normals $\tilde{\mathbf{n}}$ is:
\vspace{-3pt}
\begin{equation}
\sigma_{\mathbf{n}}^2 = 
\frac{1}{3n_r} 
\mathrm{E} \left[\|\mathbf{n}^o-\tilde{\mathbf{n}}\|^2\right].
\label{eq:noise_var_normals}
\vspace{-3pt}
\end{equation}
Obviously, we observe only noise-corrupted $\tilde{\mathbf{n}}$, and not ground truth $\mathbf{n}^o$.
However, we can first denoise $\tilde{\mathbf{n}}$ with filter $f(\,)$ and approximate $\mathbf{n}^o \approx f(\tilde{\mathbf{n}})$. 
We can thus compute the \textit{empirical} average noise variance as:
\vspace{-3pt}
\begin{equation}
\sigma_{\mathbf{n}}^2 = 
\frac{1}{3 n_r} \left(f(\tilde{\mathbf{n}}) - \tilde{\mathbf{n}}\right)^{\top} \left(f(\tilde{\mathbf{n}}) - \tilde{\mathbf{n}}\right)
\label{eq:empSNNV}
\vspace{-0pt}
\end{equation}
We derive the appropriate filter $f(\,)$ next.

Since $\mathbf{w}$ is a Gaussian random vector and $\mathbf{A}$ is a constant matrix, $(\mathbf{n}^{o}-\tilde{\mathbf{n}})$  is also a Gaussian random vector (see (\ref{eq:normal_vector_noise}) and (\ref{eq:normal_vector})). 
Thus we can design a filter $f(\,)$ by formulating an optimization problem to denoise $\mathbf{n}$ as:
\vspace{-3pt}
\begin{equation}
\min_{\mathbf{n}} \| \mathbf{n} - \tilde{\mathbf{n}} \|_2^2 + \gamma|\vert\mathbf{N}|\vert_{\text{RGLR}},
\label{eq:denoiseN}
\vspace{-3pt}
\end{equation}
where $\| \mathbf{n} - \tilde{\mathbf{n}} \|_2^2$ is the fidelity term. 

Because of our explicit search for flat patches, in this case we have a strong signal prior that $\mathbf{n}$ contains the \textit{same} normal denoted by $\mathbf{s}~=~[s_x\hspace{5pt} s_y\hspace{5pt} s_z]^{\top}$ for all points in the patch, and hence $|\vert\mathbf{N}|\vert_{\text{RGLR}}~=~0$.
Then the solution\footnote{When solving~(\ref{eq:denoiseN}), $\mathbf{s}$ should technically be constrained to have unit norm. However, this would result in a non-convex optimization problem. 
Hence we use a simple strategy where we first solve (\ref{eq:denoiseN}) without the unit-norm constraint and then normalize the solution afterwards.} to (\ref{eq:denoiseN}) defaults to the \textit{mean filter}. 
Specifically,
\vspace{-0pt}
\begin{align}
\mathbf{n}^* & = \arg \min_{s_x, s_y, s_z} 
\| \mathbf{I}_{3n_r \times 3}
\left[ \begin{array}{c}
s_x \\
s_y \\
s_z
\end{array} \right] - \tilde{\mathbf{n}} \|_2^2 
\end{align}
where $\mathbf{I}_{3n_r \times 3}$ is a column concatenation of $n_r$ $3 \times 3$ identity matrices.
Taking the derivative with respect to each of $s_x$, $s_y$ and $s_z$, we get:
\vspace{-7pt}
\begin{align}
s_x^* = \frac{1}{n_{r}}\sum_{i=1}^{n_r} \tilde{n}_{i,1}, ~~
s_y^* = \!\!\! \frac{1}{n_r}\sum_{i=1}^{n_r} \tilde{n}_{i,2}, ~~
s_z^* = \!\!\! \frac{1}{n_r}\sum_{i=1}^{n_r} \tilde{n}_{i,3}
\end{align}
where $\tilde{n}_{i,1}$, $\tilde{n}_{i,2}$ and $\tilde{n}_{i,3}$ are the $x$-, $y$- and $z$-coordinates of $\tilde{\mathbf{n}_i}$. Then after normalizing $[s_x^*\hspace{5pt} s_y^*\hspace{5pt} s_z^*]^{\top}$, we can approximate $\mathbf{n}^o\approx f(\tilde{\mathbf{n}})\approx[{\mathbf{s}^*}^{\top} \hdots {\mathbf{s}^*}^{\top}]^{\top}$, where $\mathbf{s}^{*} ~=~ [s_x^* \; s_y^* \; s_z^*]^{\top}/\sqrt{{s_{x}^{*}}^{2}+{s_{y}^{*}}^{2}+{s_{z}^{*}}^{2}}$. We can now compute the empirical average noise variance $\sigma_{\mathbf{n}}^2$ for normals using (\ref{eq:empSNNV}).

\subsubsection{Deriving $\sigma^2$ from $\sigma_{\mathbf{n}}^2$}

Having computed $\sigma_{\mathbf{n}}^2$, we now derive the relationship between $\sigma_{\mathbf{n}}^2$ and $\sigma^2$ as follows using (\ref{eq:normal_vector_noise}), (\ref{eq:normal_vector}) and (\ref{eq:noise_var_normals}):
\begin{equation}
\begin{split}
3n_r \sigma_{\mathbf{n}}^2 &= \mathrm{E}\left[(\tilde{\mathbf{n}}-\mathbf{n}^o)^{\top}(\tilde{\mathbf{n}}-\mathbf{n}^o)\right] = \mathrm{E}\left[(\mathbf{Aw})^{\top}\mathbf{Aw}\right]\\
&=
\mathrm{E}\left[\text{tr}\left(\mathbf{w}^{\top}\mathbf{A}^{\top}\mathbf{Aw}\right)\right] = \mathrm{E}\left[\text{tr}\left(\mathbf{ww}^{\top}\mathbf{A}^{\top}\mathbf{A}\right)\right] \\
&= \text{tr}\left(\mathrm{E}\left[\mathbf{ww}^{\top}\right]\mathbf{A}^{\top}\mathbf{A}\right) = \text{tr}\left(\sigma^2\mathbf{I}\mathbf{A}^{\top}\mathbf{A}\right) \\
&=\sigma^2\text{tr}\left(\mathbf{A}^{\top}\mathbf{A}\right) = \sigma^2 \sum_i \mathrm{tr} \left(\mathbf{A}_i^{\top} \mathbf{A}_i \right).
\vspace{-10pt}
\end{split}
\vspace{-10pt}
\end{equation}
where $\mathrm{E}[\mathbf{w}\mathbf{w}^T] = \sigma^2 \mathbf{I}$ because of the inter-sample independent assumption. Hence, $\sigma^2$ can be obtained as
\begin{equation}
\sigma^2=\frac{3n_r}{\sum_i \mathrm{tr}\left(\mathbf{A}_i^{\top}\mathbf{A}_i\right)}\sigma_{\mathbf{n}}^2.
\label{eq:sigmaSigma}
\end{equation}

For more reliable estimation for average noise variance $\sigma^2$, we compute it for both red and blue node sets for each flat patch, and take the average over all identified flat patches in the given point cloud. 
\vspace{-10pt}
\subsection{Laplacian Noise Variance Estimate}
\subsubsection{Optimization Formulation}

If the additive noise is Laplacian, the optimization for denoising red points in a given flat patch surface becomes:
\vspace{-8pt}
\begin{equation}
\min_{\mathbf{p}}\sum_{i=1}^{n_{r}}|\vert\mathbf{q}_i-\mathbf{p}_i\vert|_{1} + \gamma|\vert\mathbf{N}|\vert_{\text{RGLR}}
\label{eq:main_opt_laplace}
\end{equation}
Similar to (\ref{eq:denoiseN}), we also assume a strong prior that all points in the given patch have the same normals $\mathbf{s}~=~[s_x\hspace{5pt} s_y\hspace{5pt} s_z]^{\top}$,
hence $|\vert\mathbf{N}|\vert_{\text{RGLR}}~=~ 0$. 
Then, the solution to the optimization problem in (\ref{eq:main_opt_laplace}) becomes: 
\vspace{-5pt}
\begin{equation}
\min_{\mathbf{p}}\sum_{i=1}^{n_{r}}|\vert\mathbf{q}_i-\mathbf{p}_i\vert|_{1}, \hspace{10pt}\text{s.t.} \hspace{10pt} \mathbf{s} \approx\mathbf{A}_i\mathbf{p}_i+\mathbf{b}_i.  
\label{eq:lapl_opt}
\end{equation}
By solving~(\ref{eq:lapl_opt}), we find the denoised red points of the given flat patch and obtain the noise variance via the two following steps.   

\subsubsection{Upper Bound and Minimization}
\label{sec:upperbound}
Because (\ref{eq:lapl_opt}) is hard to optimize expediently, we minimize instead an upper bound of the cost function. 
Essentially, for each $\| \mathbf{q}_i - \mathbf{p}_i \|_1$,  we replace it with an upper bound $\eta\| \mathbf{A}_i \mathbf{q}_i - \mathbf{A}_i \mathbf{p}_i \|_1$ and minimize it instead, where $\eta$ is a real positive scalar. 
To achieve this goal we first derive a set of real orthonormal vectors from the eigenvectors of $\mathbf{A}_i$ to span the $\mathbb{R}^3$ space, so that $\mathbf{q}_i$ and $\mathbf{p}_i$ can be expressed in terms of them.

Since $\mathbf{A}_i$ is a $3\times 3$ \textit{skew symmetric} matrix (\textit{i.e.}, $\mathbf{A}_i^{\top}~=~-\mathbf{A}_i$), its three eigenvalues\footnote{See the properties of skew symmetric matrix at https://en.wikipedia.org/wiki/Skew-symmetric\_matrix} are $0$, $i\lambda$, and $-i\lambda$, where $\lambda$ is a real positive scaler and the corresponding orthonormal\footnote{Since $\mathbf{A}_i$ is a \textit{normal} matrix (\textit{i.e.} $\mathbf{A}_i^{\top}\mathbf{A}_i~=~\mathbf{A}_i\mathbf{A}_i^{\top}$), its eigenvectors are orthonormal.} eigenvectors are $\mathbf{u}_1$, $\mathbf{u}_2$, $\mathbf{u}_3$ respectively, where $\mathbf{u}_1$ is a real eigenvector. 
Further, $\mathbf{u}_2$ and $\mathbf{u}_3$ are complex conjugate of each other (the proof is provided in Section 1.B in the supplement). 
Being complex conjugates, $\beta_1(\mathbf{u}_2+\mathbf{u}_3)$ (denoted by $\mathbf{v}_2$) and $i\beta_2(\mathbf{u}_2-\mathbf{u}_3)$ (denoted by $\mathbf{v}_3$) are real vectors and are orthonormal, where $\beta_1$ and $\beta_2$ are corresponding normalization factors. Then $\mathbf{v}_1~=~\mathbf{u}_1$, $
\mathbf{v}_2$ and $\mathbf{v}_3$ are three orthonormal basis vectors that span the $\mathbb{R}^3$ space, and we can write $\mathbf{p}_i$ and $\mathbf{q}_i$ as a linear combination of $\mathbf{v}_1$, $\mathbf{v}_2$, and $\mathbf{v}_3$. Specifically,
\vspace{-5pt}
\begin{equation}
\mathbf{q}_i=\sum_{j=1}^3\hat{q}_{ij}\mathbf{v}_j, \hspace{10pt} \mathbf{p}_i=\sum_{j=1}^3\hat{p}_{ij}\mathbf{v}_j, 
\label{eq:basis}
\vspace{-2pt}
\end{equation}
where $\hat{q}_{i,j}$ and $\hat{p}_{i,j}$ (for $j=1,2,3$) are real values.

Next, using~(\ref{eq:basis}), we derive an upper bound $\eta\| \mathbf{A}_i \mathbf{q}_i-\mathbf{A}_i \mathbf{p}_i \|_1$ for $\| \mathbf{q}_i - \mathbf{p}_i \|_1$. 
First, to avoid mapping any energy of $\mathbf{q}_i-\mathbf{p}_i$ into the null space of $\mathbf{A}_i$ (\textit{i.e.} space spanned by $\mathbf{v}_1$), we set $\hat{p}_{i1}~=~\hat{q}_{i1}$. Then by substituting $\mathbf{v}_2~=~\beta_1(\mathbf{u}_2+\mathbf{u}_3)$ and $\mathbf{v}_3~=~i\beta_2(\mathbf{u}_2-\mathbf{u}_3)$, we have,
\vspace{-6pt}
\begin{equation}
\begin{split}
\mathbf{q}_i-\mathbf{p}_i=\tilde{u}_2\mathbf{u}_2+\tilde{u}_3\mathbf{u}_3,
\end{split}
\label{eq:newv}
\vspace{-15pt}
\end{equation}
where $\tilde{u}_2~=~\beta_1(\hat{q}_{i2}-\hat{p}_{i2})+i\beta_2(\hat{q}_{i3}-\hat{p}_{i3})$ and $\tilde{u}_2~=~\beta_1(\hat{q}_{i2}-\hat{p}_{i2})+i\beta_2(\hat{p}_{i3}-\hat{q}_{i3})$. 
According to the orthogonal transformation\footnote{Orthogonal transformation of $\mathbf{A}_i$ is given as $\mathbf{A}_i~=~\begin{bmatrix}
\mathbf{u}_1 & \mathbf{u}_2 & \mathbf{u}_3
\end{bmatrix}\text{diag}(0, i\lambda, -i\lambda)\begin{bmatrix}
\mathbf{u}_1 & \mathbf{u}_2 & \mathbf{u}_3
\end{bmatrix}^\text{H}.$}, $\mathbf{A}_i~=~i\lambda\mathbf{u}_2\mathbf{u}_2^\text{H}-i\lambda\mathbf{u}_3\mathbf{u}_3^\text{H}$ and then we can write
\vspace{-0pt}
\begin{equation}
\norm{\mathbf{A}_i(\tilde{u}_2\mathbf{u}_2+\tilde{u}_3\mathbf{u}_3)}_1=\lambda\norm{\tilde{u}_2\mathbf{u}_2-\tilde{u}_3\mathbf{u}_3}_1.
\label{eq:mulwithmat}
\vspace{-0pt}
\end{equation}

Further, for any vector $\mathbf{z}\in\mathbb{R}^3$ with $\norm{\mathbf{z}}_2~=~r$ where $r$ is a constant, we can prove that $r~\leq~\norm{\mathbf{z}}_1~\leq~\sqrt{3}r$ (the proof is in Section 1.C of the supplement).
  Moreover, since $\mathbf{u}_2$ and $\mathbf{u}_3$ are orthogonal, one can see that $\norm{\tilde{u}_2\mathbf{u}_2+\tilde{u}_3\mathbf{u}_3}_2~=~\norm{\tilde{u}_2\mathbf{u}_2-\tilde{u}_3\mathbf{u}_3}_2~=~c$. Therefore, by the proof in the supplement, we have $c~\leq~ \norm{\tilde{u}_2\mathbf{u}_2+\tilde{u}_3\mathbf{u}_3}_1~\leq~ \sqrt{3}c$ and $c~\leq~\norm{\tilde{u}_2\mathbf{u}_2-\tilde{u}_3\mathbf{u}_3}_1~\leq~\sqrt{3}c$.
We can write
\vspace{0pt}
\begin{equation}
\norm{\tilde{u}_2\mathbf{u}_2+\tilde{u}_3\mathbf{u}_3}_1\leq \lambda\norm{\tilde{u}_2\mathbf{u}_2-\tilde{u}_3\mathbf{u}_3}_1, \hspace{8pt} \text{if} \hspace{8pt} \lambda\geq\sqrt{3}.
\label{eq:upper}
\vspace{-0pt}
\end{equation}
In this case we cannot guarantee that for all $\mathbf{A}_i$, $\lambda\geq\sqrt{3}$. Here we multiply each $\mathbf{A}_i$ with a common real positive scalar $\eta$ so that the corresponding $\lambda\geq \sqrt{3}$. Therefore, we can select $\eta~=~\sqrt{3}/\chi$, where $\chi$ is the minimum $\lambda$ among all $\mathbf{A}_i$.
By combining (\ref{eq:upper}) and (\ref{eq:mulwithmat}) with (\ref{eq:newv}), we have
\vspace{-2pt}
\begin{equation}
|\vert\mathbf{q}_i-\mathbf{p}_i\vert|_{1}\leq\eta\norm{\mathbf{A}_i{\mathbf{q}_i}-\mathbf{A}_i{\mathbf{p}_i}}_1,
\label{eq:upperbound}
\end{equation}
Using (\ref{eq:upperbound}), we can write an upper bound for (\ref{eq:lapl_opt}) as follows:
\vspace{-2pt}
\begin{equation}
\begin{split}
&\min_{\mathbf{p}}\sum_{i=1}^{n_r}|\vert\mathbf{q}_i-\mathbf{p}_i\vert|_{1}\leq \min_{\mathbf{p}}\sum_{i=1}^{n_r}\eta|\vert\mathbf{A}_i\mathbf{q}_i-\mathbf{A}_i\mathbf{p}_i\vert|_{1} \\
& \hspace{10pt}\text{s.t.} \hspace{10pt} \mathbf{s} \approx\mathbf{A}_i\mathbf{p}_i+\mathbf{b}_i.  
\end{split}
\label{eq:lapl_opt_upper}
\vspace{-2pt}
\end{equation}
Here for simplicity, instead of minimizing (\ref{eq:lapl_opt}), we minimize its upper bound in (\ref{eq:lapl_opt_upper}) by substituting the constraint $\mathbf{s} ~\approx~\mathbf{A}_i\mathbf{p}_i+\mathbf{b}_i$ to the cost function as follow:
\vspace{-5pt}
\begin{equation}
\min_{\mathbf{s}}\sum_{i=1}^{n_r}|\vert\mathbf{A}_i\mathbf{q}_i-\mathbf{b}_i - \mathbf{s}\vert|_{1}.
\label{eq: Lapopt}
\vspace{-5pt}
\end{equation}

It is known\footnote{http://web.uvic.ca/~dgiles/blog/median2.pdf} that the median filter minimizes the $\ell_1$-norm with respect to a set of scalars. Hence, the solution\footnote{Following the similar strategy used in (\ref{eq:denoiseN}), we solve (\ref{eq: Lapopt}) without the normalization constraint for $\mathbf{s}$ and then normalize the solution afterwards.} to~(\ref{eq: Lapopt}) defaults to the \textit{median filter} as follow:
\begin{equation}
s_x^* = \mathrm{med}\left(\{{a}^i_x\}\right),~
s_y^* = \mathrm{med}\left(\{{a}^i_y\}\right),~
s_z^* = \mathrm{med}\left(\{{a}^i_z\}\right)
\end{equation}
where $\mathrm{med}(\mathcal{S})$ computes the median of the set of scalars in set $\mathcal{S}$, and ${a}^i_x$, ${a}^i_y$, ${a}^i_x$ are $x, y, z$ entries of $\mathbf{A}_i\mathbf{q}_i-\mathbf{b}_i$. Finally, $\mathbf{s}$ is computed as the normalized $[s_x^*\hspace{5pt} s_y^*\hspace{5pt} s_z^*]^{\top}$.

\subsubsection{Computing Noise Variance $\sigma^2$ from Normal $\mathbf{s}$}

After computing $\mathbf{s}$, we can obtain the following two equations based on the linear relationship of $\mathbf{s} ~\approx~\mathbf{A}_i\mathbf{p}_i+\mathbf{b}_i$.
\begin{equation}
\mathbf{v}_2^{\top}(\mathbf{s}-\mathbf{b}_i)=\mathbf{v}_2^{\top}\mathbf{A}_i\mathbf{p}_i,\hspace{5pt} \mathbf{v}_3^{\top}(\mathbf{s}-\mathbf{b}_i)=\mathbf{v}_3^{\top}\mathbf{A}_i\mathbf{p}_i.
\label{eq:p_find}
\end{equation}
Moreover, $\hat{p}_{i1}~=~\mathbf{v}_1^{\top}\mathbf{p}_i$, $\hat{q}_{i1}~=~\mathbf{v}_1^{\top}\mathbf{q}_i$, and we set $\hat{p}_{i1}~=~\hat{q}_{i1}$ at Section~\ref{sec:upperbound}. Now using this relationship and (\ref{eq:p_find}), we can obtain $\mathbf{p}_i$ by solving the following linear equations,
\begin{equation}
\underbrace{\begin{bmatrix}
\mathbf{v}_1^{\top}\\
\mathbf{v}_2^{\top}\mathbf{A}_i\\
\mathbf{v}_3^{\top}\mathbf{A}_i
\end{bmatrix}}_{\mathbf{A}_{{v}}}\mathbf{p}_i=\begin{bmatrix}
\mathbf{v}_1^{\top}\mathbf{q}_i\\
\mathbf{v}_2^{\top}(\mathbf{s}-\mathbf{b}_i)\\
\mathbf{v}_3^{\top}(\mathbf{s}-\mathbf{b}_i)
\end{bmatrix}.
\label{eq:p_sol}
\end{equation}
In order to solve (\ref{eq:p_sol}) for $\mathbf{p}_i$, $\mathbf{A}_v$ should be invertible. We prove that $\mathbf{A}_v$ is invertible in Section 1.D of the supplement. 
After estimating $\mathbf{p}_i$, we can compute the empirical average noise variance of the point cloud using the red points in the given flat patch as follow:
\vspace{-7pt}
\begin{equation}
\sigma^2=\frac{1}{3n_r}\sum_{i=1}^{n_r}(\mathbf{q}_i-\mathbf{p}_i)^{\top}(\mathbf{q}_i-\mathbf{p}_i).
\end{equation}
For more reliable estimation for $\sigma^2$, we compute it for both red and blue node sets for each flat patch, and take the average over all identified flat patches in the given point cloud.

\vspace{-10pt}
\subsection{Discussion}
\label{sec:dis}
One can see the simplicity of our proposed variance estimation method that involves only: i) the mean filters for $\ell_2$-norm fidelity term, and ii) median filters, eigen-decomposition and matrix inversion of $3\times 3$ matrices for $\ell_1$-norm fidelity term. 
The bipartite graph approximation and computation of the surface normals for the red and blue nodes are more costly as described in Section~\ref{sec:problem}, but this is performed necessarily as part of the denoising algorithm. 
Given that, the marginal computation overhead is small.

An alternative approach is to perform \textit{principal component analysis} (PCA)~\cite{abdi2010} directly on the observed 3D points in a flat patch to obtain a best-fitted 2D plane in a $\ell_2$-norm sense, then compute the Euclidean distances between the observed points and the plane to estimate noise variance $\sigma^2$.
However, PCA is known to be fragile to sparse but large noise~\cite{mattei2017}. 
Hence, in the case of sparse Laplacian-like noise, PCA-based approach\footnote{There exist extensive works on \textit{robust PCA} (RPCA)~\cite{candes2011} that can handle large sparse noise, but these methods require iterative minimization of the nuclear norm, solved via \textit{singular value decomposition} (SVD) of complexity $O(n^3)$.} would not perform well. 
We demonstrate this point empirically through experiments in Section\;\ref{sec:exp}.

%% file: results.tex
We present comprehensive experiments to verify the effectiveness of the proposed algorithms in solving the point cloud denoising problem. 
There are three sets of results: i) weight parameter computation based on estimated noise variance, ii) noise variance estimation, and iii) denoising. 
All the experiments consider both non-sparse noise like Gaussian and sparse noise like Laplacian. Further, all the point cloud models are first rescaled, so that each tightly fits inside a bounding box with the same diagonal.

For numerical comparisons of denoising results, we measure the point-to-point (C2C) error and point to plane (C2P) error~\cite{tian2017} between ground truth and denoising points sets. 
For C2C error, we first measure the average of the squared Euclidean distances between ground truth points and their closest denoised points, and also that between the denoised points and their closest ground truth points. Then the smaller between these two measures is taken as C2C. In C2P error, we first measure the average of the squared Euclidean distances between ground truth points and tangent planes at their closest denoised points, and also that between the denoised points and tangent planes at their closest ground truth points. Then the smaller between these two measures is taken as the C2P.
\\
\textbf{Results for Weight Parameter Estimation:} 
In order to validate the linear relationship between $\gamma_{opt}$ and $\sigma^2$ (as stated in~(\ref{eq:relation_gam_sig})) for Gaussian noise and Laplacian noise,
we use six point cloud models from~\cite{levoy2005}, four models from~\cite{Microsoft2016}, and ten models from~\cite{nouri2017technical}.
First, Gaussian / Laplacian noise with zero mean and standard deviation $\sigma$ of 0.1, 0.15, $\hdots$, 0.5 is added to the 3D position of these clean point clouds. 
Then we denoise point cloud models with Gaussian noise following Section~\ref{sec:sol for l1} and models with Laplacian noise following Section~\ref{sec:l2fid} for different $\gamma$ values ($\gamma$ from 0 to $1$ with interval 0.01). 
For a given point cloud and a given noise level, $\gamma_{opt}$ is chosen as the $\gamma$ value that gives the minimum C2P between the denoised point cloud and its ground truth. 
(C2P is a metric to measure the denoising quality, which we will discuss soon). 

\begin{figure}[t]
\centering
\hspace{-3pt}
\subfloat[for Gaussian noise]{
\includegraphics[width=0.24\textwidth]{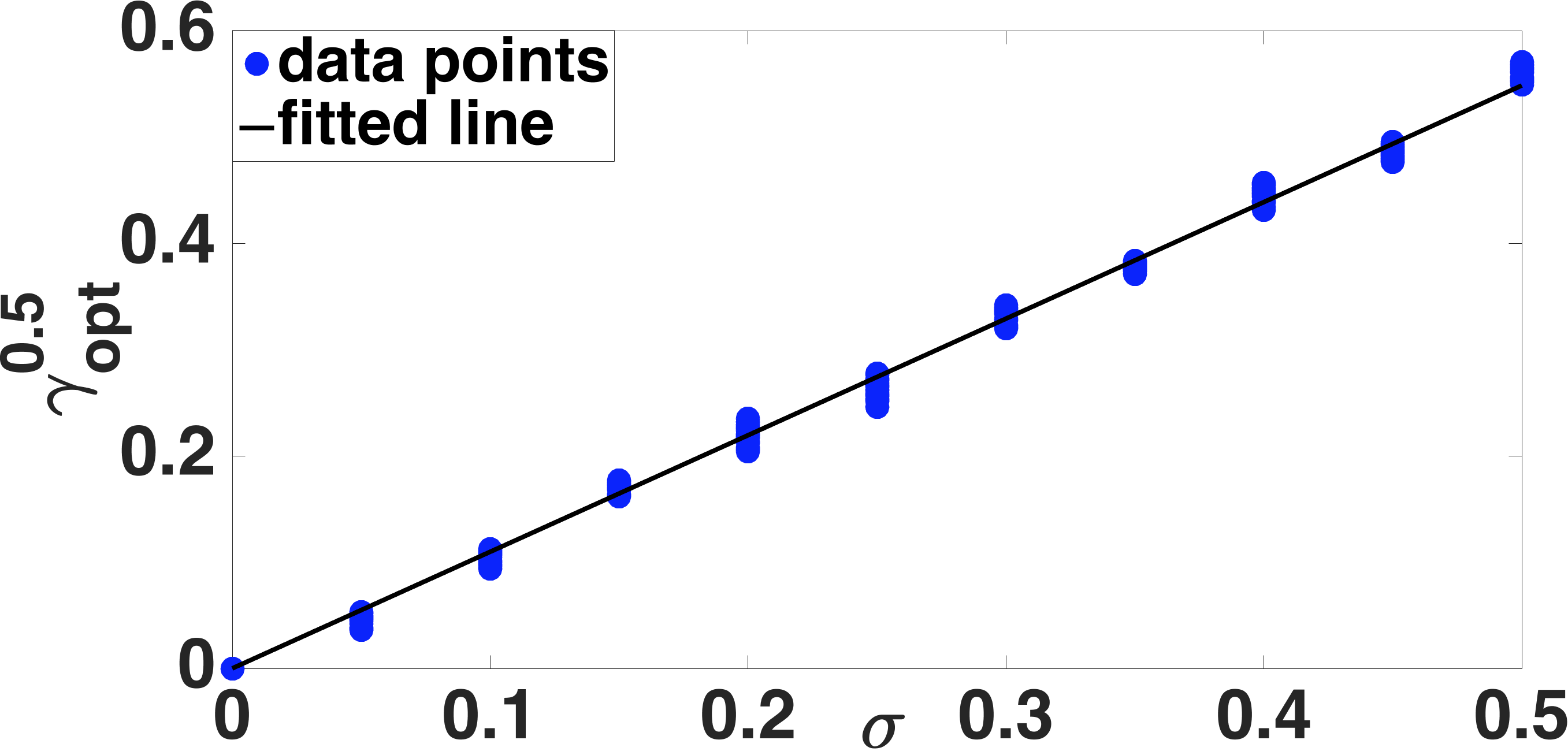}
}
\subfloat[for Laplacian noise]{
\includegraphics[width=0.235\textwidth]{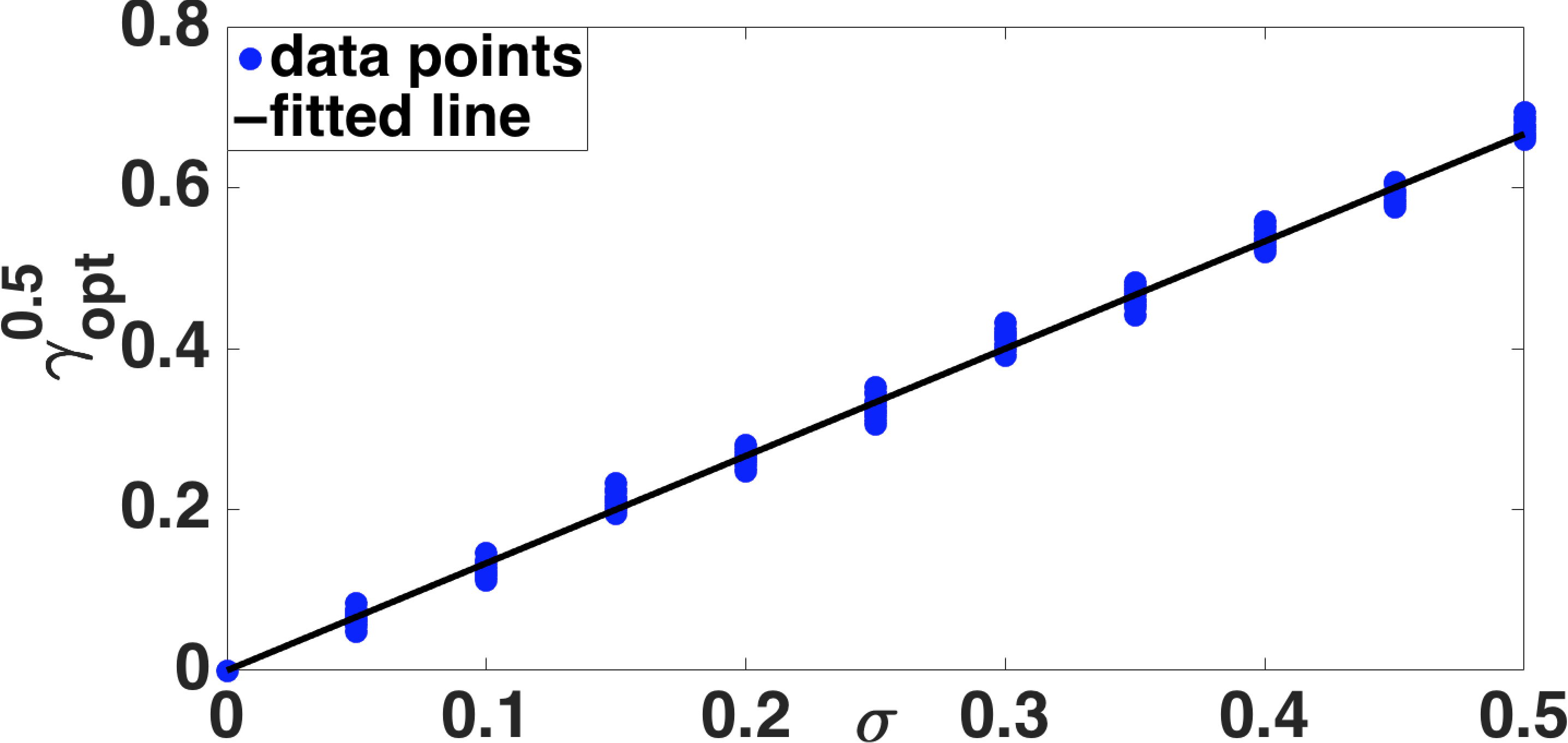}
}
\vspace{-5pt}
\caption[]{$\sigma$ vs $\sqrt{\gamma_{opt}}$ (a line is fitted based on minimum mean squire error).} 
\label{fig:opt_gamma}
\vspace{-17pt}
\end{figure}

We plot $\sigma$ vs $\sqrt{\gamma_{opt}}$, averaged over different point cloud models, in Fig.~\ref{fig:opt_gamma}.
We observe that there is an approximate linear relationship between $\gamma_{opt}$ and $\sigma^2$ as in~(\ref{eq:relation_gam_sig}). 
We can thus determine constant $\alpha$ in (\ref{eq:relation_gam_sig}) empirically using Fig.~\ref{fig:opt_gamma}. 
It should be noted that finding $\alpha$ is an off-line process and then we can determine $\gamma_{opt}$ for a given noisy point cloud based on its estimated noise variance.\\
\textbf{Results for Noise Variance Estimation:} The proposed noise variance estimation method is compared with the PCA-based method mentioned in Section~\ref{sec:dis}. 
Point cloud models we use are Bunny provided in~\cite{levoy2005}, Gargoyle, DC, and Daratech provided in \cite{rosman2013,mattei2017}. 
To show the performance of noise variance estimation quantitatively, the relative percentage error ($\epsilon$) in terms of the noise standard deviation (SD) is computed as follows:
\vspace{-3pt}
\begin{equation}
\epsilon=\frac{|\sigma-\hat{\sigma}|}{\sigma}\times 100\%,
\vspace{-3pt}
\end{equation}
where $\hat{\sigma}$ is the SD of the estimated noise and $\sigma$ represents the SD of the actual noise. 

Gaussian / Laplacian noise with zero mean and standard deviation $\sigma$ of 0.1, 0.2, $\hdots$, 0.5 is added to the 3D position of the clean point clouds. The estimated noise variance (in terms of $\sigma_e$) and the relative percentage error  are given in Table~\ref{tab:noisevar_gau} (for Gaussian noise) and Table~\ref{tab:noisevar_lap} (for Laplacian noise). On average, we observe that our estimation and PCA-based estimation for Gaussian noise have approximately similar performance. However, for Laplacian noise, the proposed method is more accurate than PCA-based scheme, with $\epsilon$ reduced by 43\% on average.\\
\textbf{Denoising Results:} The proposed point cloud denoising method is compared with four existing methods: APSS \cite{guennebaud2007}, RIMLS~\cite{oztireli2009}, AWLOP\cite{huang2013}, the moving robust principle component analysis (MRPCA) algorithm~\cite{mattei2017}, and our recently proposed GTV-based method~\cite{dinesh2018fast}. APSS and RIMLS are implemented with MeshLab software~\cite{cignoni2008}, AWLOP is implemented with EAR software~\cite{huang2013}, and the source code of MRPCA is provided by the authors. 
Point cloud models we use are Bunny provided in~\cite{levoy2005}, Gargoyle, DC, Daratech, Anchor, Lordquas, Fandisk, and Laurana provided in \cite{rosman2013,mattei2017}. 
We note that these models are different from the models that we used to determine $\alpha$ in~(\ref{eq:relation_gam_sig}). 
Both quantitative and visual comparisons are presented.


\vspace{-5pt}
\subsubsection{Gaussian Noise}
Gaussian noise with zero mean and standard deviation 
of 0.2 and 0.4 is added to the 3D coordinates of the point cloud.
Quantitative results in terms of C2C and C2P errors are shown in Table~\ref{tab:gauusian_0.2} (for $\sigma~=~0.2$) and Table~\ref{tab:gauusian_0.4} (for $\sigma~=~0.4$), where the proposed
method is shown to have the lowest C2C and C2P errors. For Gaussian noise, the proposed denoising method (\textit{i.e.} in Section~\ref{sec:sol for l1}) has two different approaches of solving~(\ref{eq:sol_qud}):  CG and GSF. 
The quantitative results show that the both approaches have approximately similar performance in terms of C2C and C2P. As shown in Table~\ref{tab:gauusian_0.2} and~\ref{tab:gauusian_0.4}, C2C and C2P of the proposed method are clearly better than the competing schemes, with C2C and C2P reduced by 15\% and 36\% (on average), respectively.  

\begin{figure*}[t]
\centering
\subfloat[ground truth]{
\includegraphics[width=0.162\textwidth]{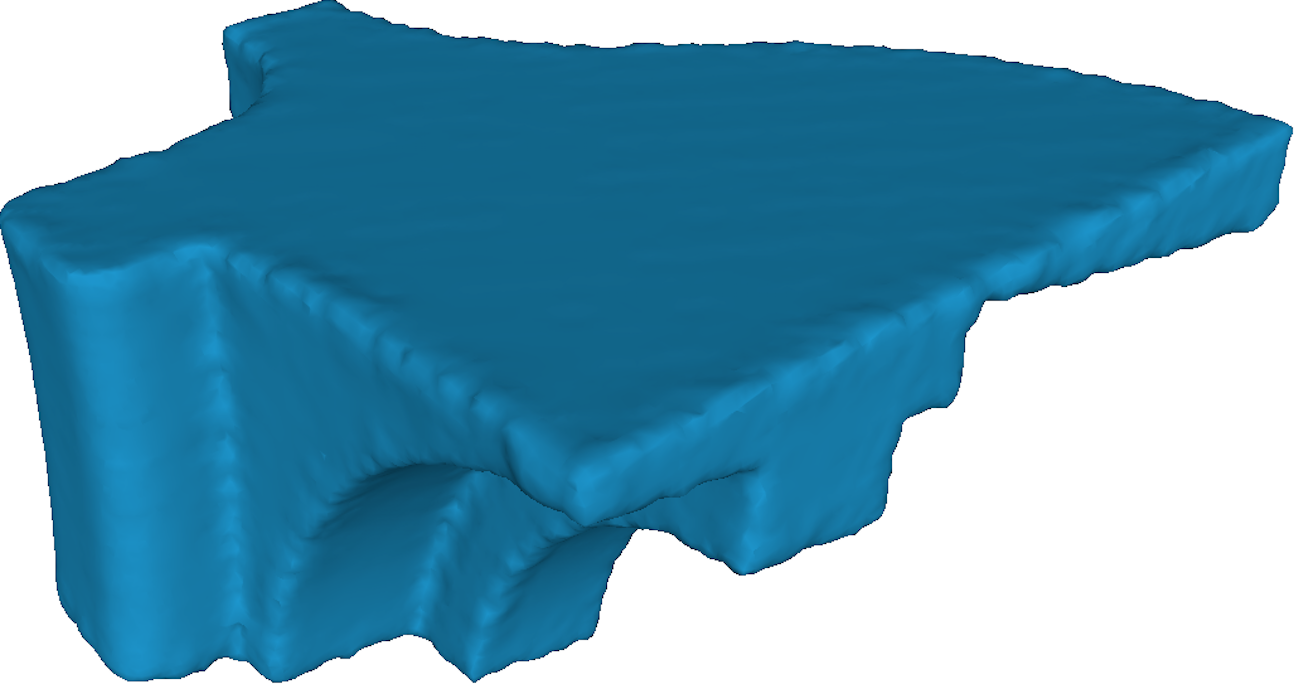}
\label{fig:pcd-day}}
\hspace{-8pt}
\subfloat[noisy input]{
\includegraphics[width=0.162\textwidth]{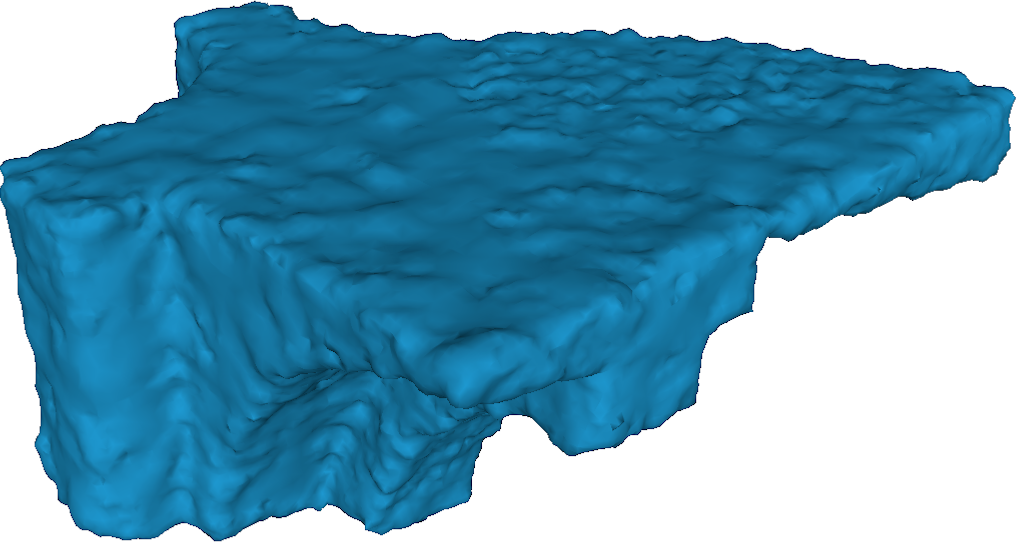}
\label{fig:pcd-on}}
\hspace{-8pt}
\subfloat[{APSS}]{
\includegraphics[width=0.162\textwidth]{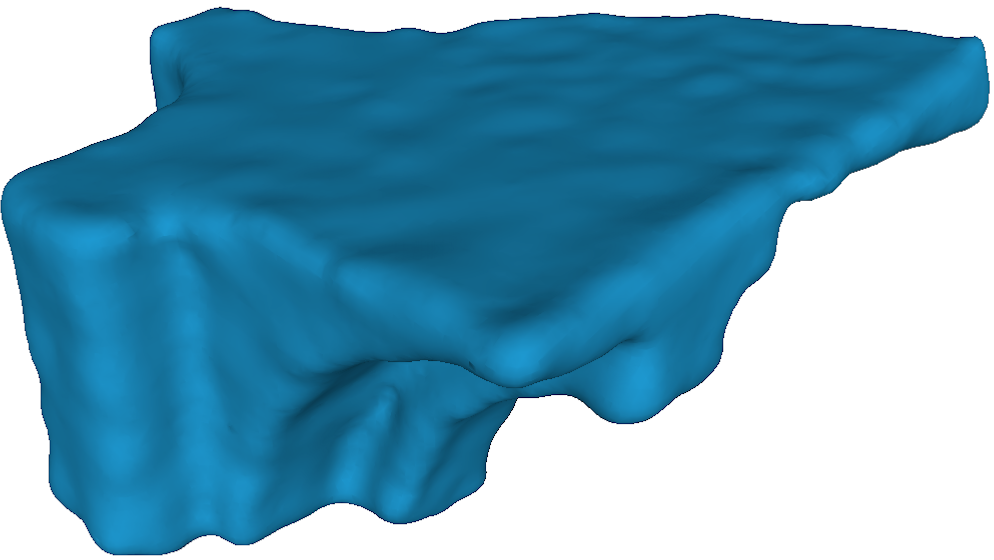}
\label{fig:pcd-off}}
\hspace{-8pt}
\subfloat[RIMLS]{
\includegraphics[width=0.162\textwidth]{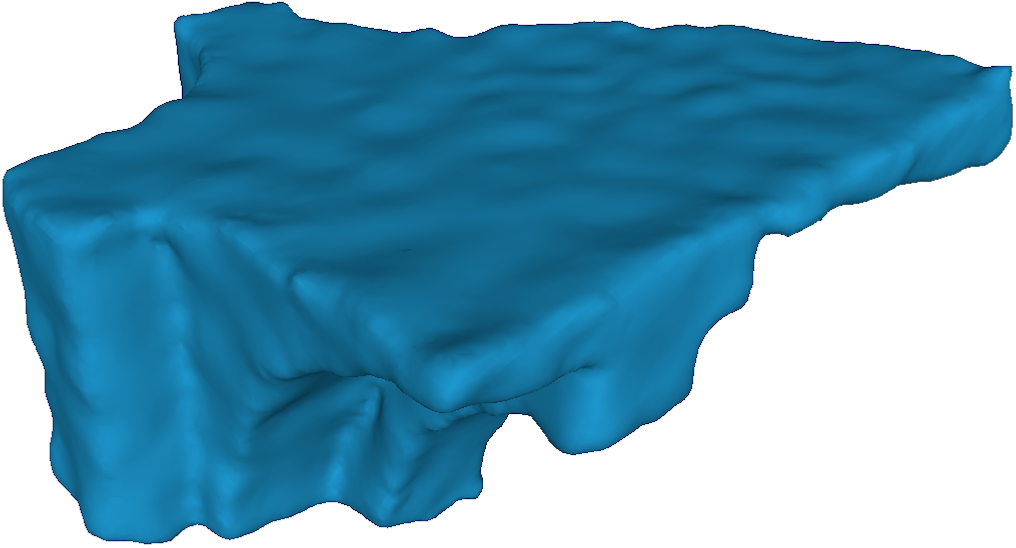}
\label{fig:pcd-off}}
\hspace{-8pt}
\subfloat[MRPCA]{
\includegraphics[width=0.162\textwidth]{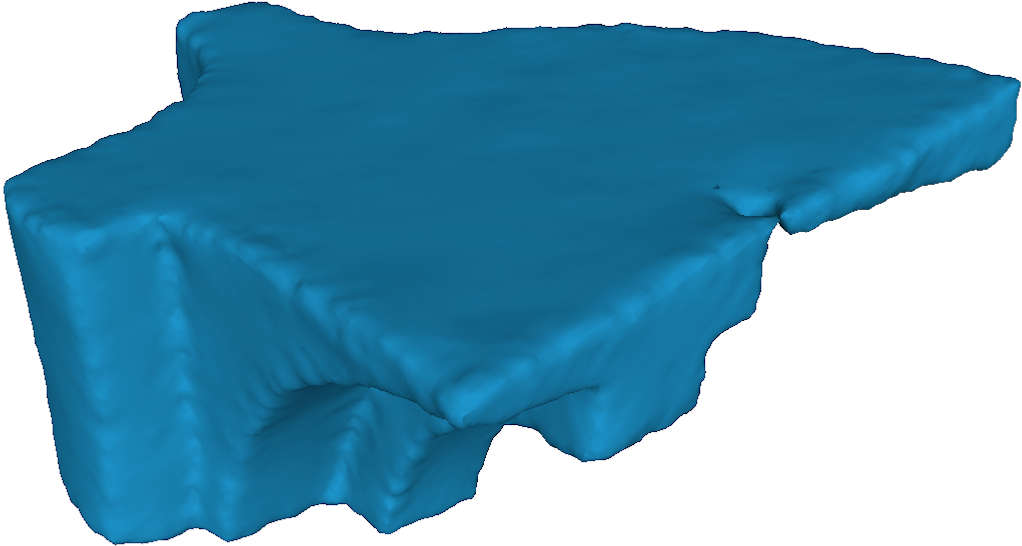}
\label{fig:tvlcd-ofn}}
\hspace{-8pt}
\subfloat[proposed]{
\includegraphics[width=0.162\textwidth]{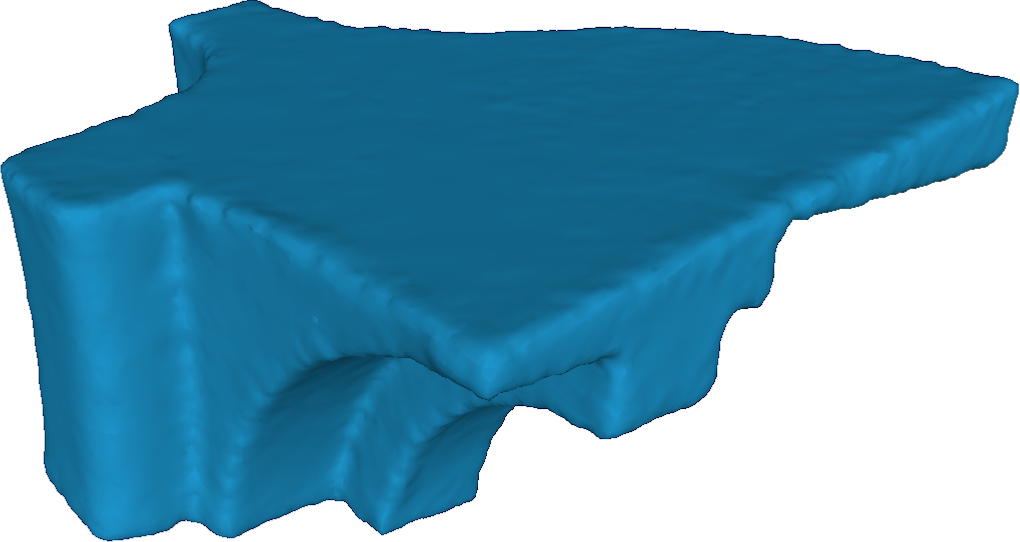}
\label{fig:tvlcd-day}}
\hspace{-8pt}
\vspace{-5pt}
\caption[Denoising results illustration for Fandisk model ($\sigma=0.4$)]{Denoising results illustration for Fandisk model (Gauusian noise with $\sigma=0.4$); a surface is fitted over the point cloud for better visualization.} 
\label{fig:fand_gau}
\vspace{-10pt}
\end{figure*}

Visual results for Fandisk model is shown in Fig.~\ref{fig:fand_gau}. The results of APSS and RIMLS schemes are over-smoothed, and the results of MRPCA scheme are distorted with some details lost.
For the proposed method, the details are well preserved without over-smoothing. Additional visual results are given in Fig. 1 and 2 in the supplement.    
\vspace{-5pt}
\subsubsection{Laplacian Noise}

Laplacian noise with zero mean and standard deviation 
of 0.1 and 0.3 is added to the 3D coordinates of the point cloud.
Quantitative results in terms of C2C and C2P errors are shown in Table~\ref{tab:laplace_0.1} (for $\sigma~=~0.1$) and Table~\ref{tab:laplace_0.3} (for $\sigma~=~0.3$). 
The proposed method (\textit{i.e.} in Section~\ref{sec:l2fid}) is shown to have the lowest C2C and C2P errors; our method is significantly better than the competing schemes, with C2C and C2P reduce by 20\% and 43\% (on average) respectively. 
To show our proposed method for $\ell_1$-norm fidelity term is more appropriate to model Laplacian noise, we present denoising results using the method in Section~\ref{sec:sol for l1} as well (in Table~~\ref{tab:laplace_0.1} and ~\ref{tab:laplace_0.3}, we have reported the best results out of GC and GSF approaches in Section~\ref{sec:sol for l1}). 
We observe that for Laplacian noise, the proposed method with $\ell_1$-norm fidelity term substantially outperforms the method of $\ell_2$-norm fidelity term, with C2C and C2P reduced by 12\% and 22\% (on average) respectively.    

\begin{figure*}[t]
\centering
\subfloat[ground truth]{
\includegraphics[width=0.162\textwidth]{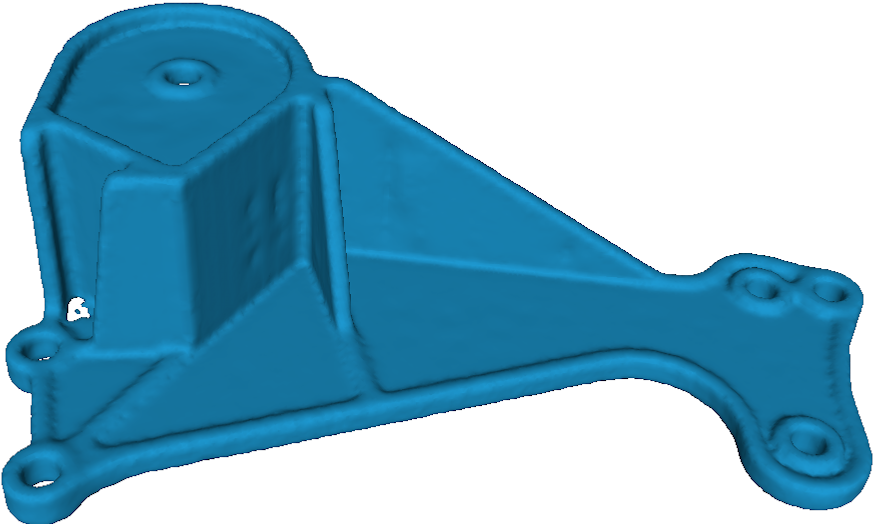}
\label{fig:pcd-day}}
\hspace{-8pt}
\subfloat[noisy input]{
\includegraphics[width=0.162\textwidth]{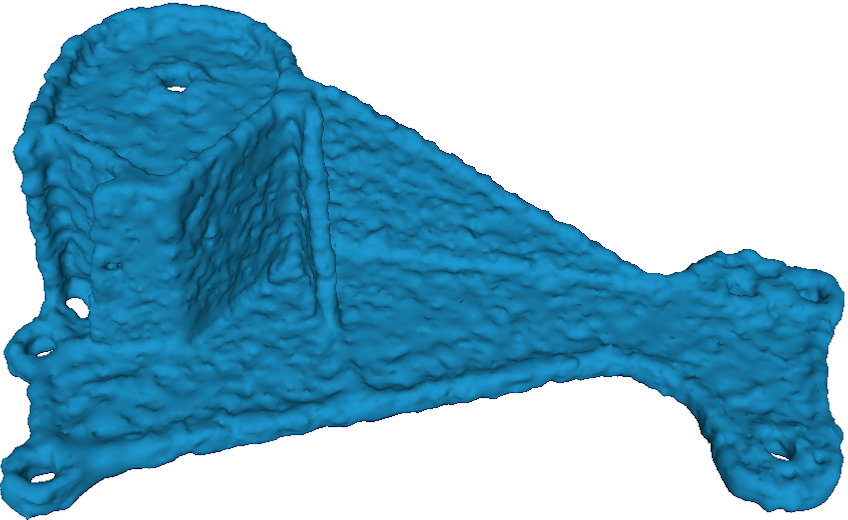}
\label{fig:pcd-on}}
\hspace{-8pt}
\subfloat[{APSS}]{
\includegraphics[width=0.162\textwidth]{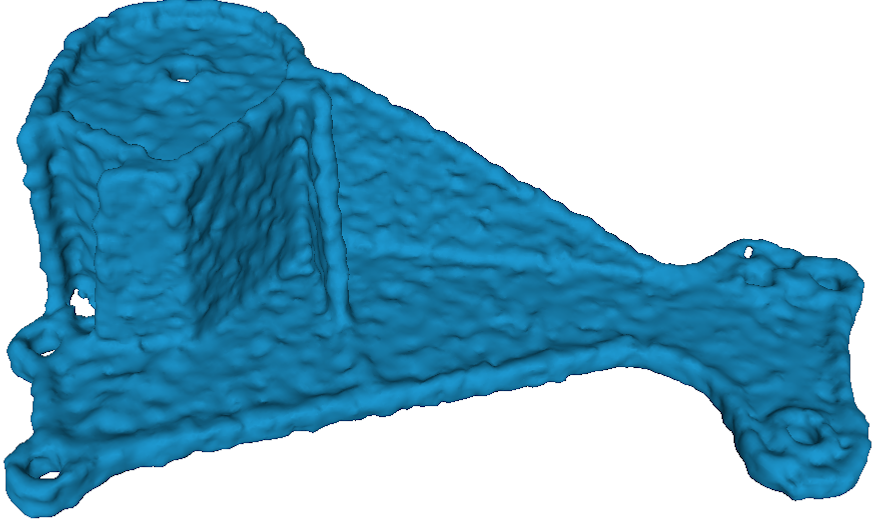}
\label{fig:pcd-off}}
\hspace{-8pt}
\subfloat[RIMLS]{
\includegraphics[width=0.162\textwidth]{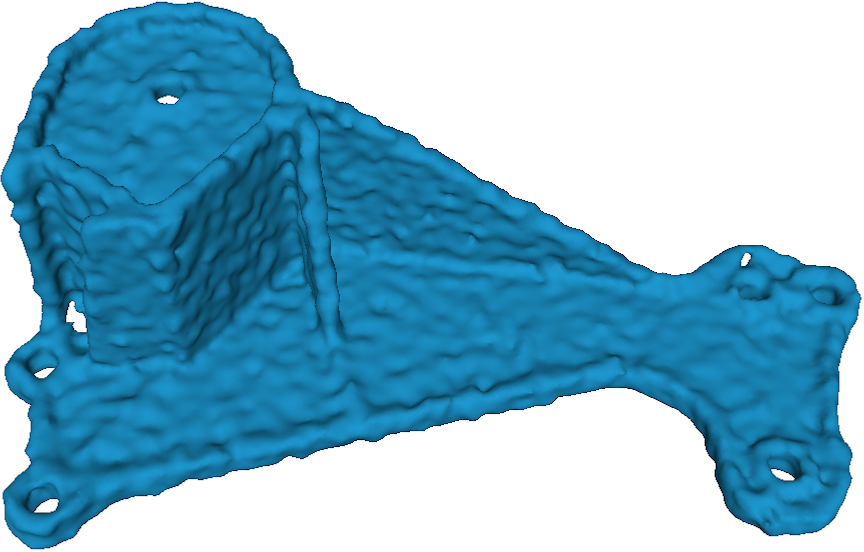}
\label{fig:pcd-off}}
\hspace{-8pt}
\subfloat[MRPCA]{
\includegraphics[width=0.162\textwidth]{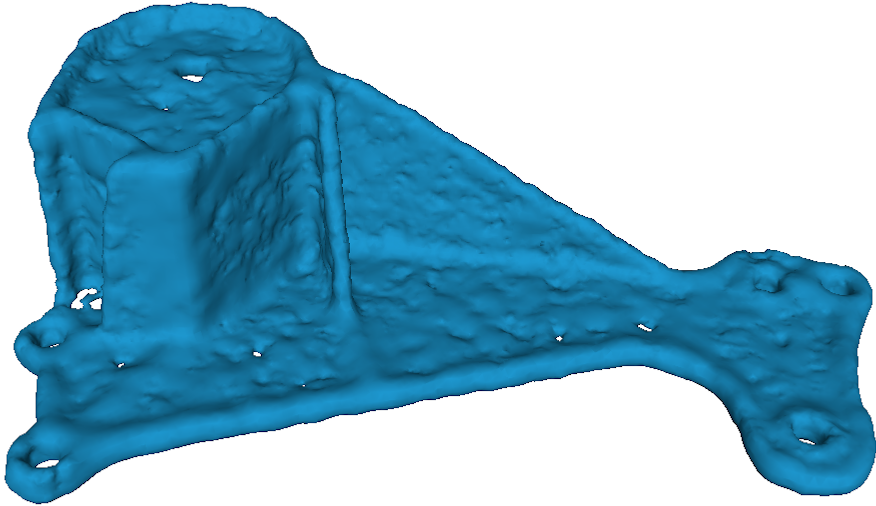}
\label{fig:tvlcd-ofn}}
\hspace{-8pt}
\subfloat[proposed]{
\includegraphics[width=0.162\textwidth]{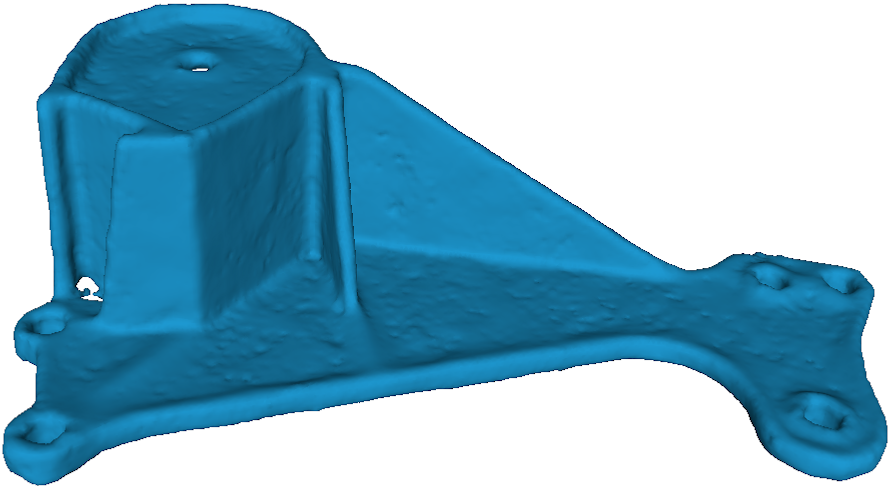}
\label{fig:tvlcd-day}}
\hspace{-8pt}
\vspace{-5pt}
\caption[Denoising results illustration for Daratech model (Laplacian noise with $\sigma=0.3$)]{Denoising results illustration for Daratech model ($\sigma=0.3$); a surface is fitted over the point cloud for better visualization.} 
\label{fig:dara_lap}
\vspace{-15pt}
\end{figure*}

Visual results for Daratech model are shown in Fig.~\ref{fig:dara_lap}. 
For the proposed method, sharp features are well preserved without over-smoothing compared to the existing methods. Additional visual results are given in Fig. 3 and 4 in the supplement.

\begin{table*}[t]
\centering
\caption{Noise Estimation Results in terms of $\hat{\sigma}$ and $\epsilon$ (\%) for Gaussian Noise}
\vspace{-10pt}
\begin{tabular}{|c|c;{0.5pt/1pt}c|c;{0.5pt/1pt}c|c;{0.5pt/1pt}c|c;{0.5pt/1pt}c|c;{0.5pt/1pt}c|c;{0.5pt/1pt}c|c;{0.5pt/1pt}c|c;{0.5pt/1pt}c|c|}
\hline
\multirow{3}{*}{\begin{tabular}[c]{@{}c@{}}Actual\\ $\sigma$\end{tabular}} & \multicolumn{16}{c|}{Estimated $\sigma$ ($\hat\sigma$) and $\epsilon$ (\%) respectively}                                                                                                                                                                                     \\ \cline{2-17} 
                                                                          & \multicolumn{4}{c|}{Bunny}                            & \multicolumn{4}{c|}{Gargoyle}                         & \multicolumn{4}{c|}{DC}                               & \multicolumn{4}{c|}{Daratech}                         \\ \cline{2-17} 
                                                                          & \multicolumn{2}{c|}{Prop.} & \multicolumn{2}{c|}{PCA} & \multicolumn{2}{c|}{Prop.} & \multicolumn{2}{c|}{PCA} & \multicolumn{2}{c|}{Prop.} & \multicolumn{2}{c|}{PCA} & \multicolumn{2}{c|}{Prop.} & \multicolumn{2}{c|}{PCA} \\ \hline
0.1                                                                      & \textbf{0.112}        & \textbf{12.0}        & 0.113       & 13.0       & \textbf{0.112}        & \textbf{12.0}        & 0.114       & 14.0       & 0.115        & 15.0        & \textbf{0.114}       & \textbf{14.0}       & 0.114        & 14.0        & \textbf{0.112}       & \textbf{12.0}       \\ \hline

0.2                                                                      & \textbf{0.211}        & \textbf{5.5}        & 0.217       & 8.5       & 0.225        & 12.5        & \textbf{0.223}       & \textbf{11.5}       & \textbf{0.184}        & \textbf{8.0}        & 0.225       & 12.5       & 0.237        & 18.5        & \textbf{0.233}       & \textbf{16.5}       \\ \hline

0.3                                                                      & \textbf{0.332}        & \textbf{10.7}        & 0.342       & 14.0       & \textbf{0.323}        & \textbf{7.7}        & 0.327       & 9.0       & \textbf{0.273}        & \textbf{9.0}        & 0.334       & 11.3       & \textbf{0.344}        & \textbf{14.7}        & 0.348       & 16.0       \\ \hline

0.4                                                                      & \textbf{0.446}        & \textbf{11.5}        & 0.451       & 12.8       & 0.445        & 11.3        & \textbf{0.443}       & \textbf{10.8}       & \textbf{0.362}        & \textbf{9.5}        & 0.453       & 13.3       & \textbf{0.446}        & \textbf{11.5}        & 0.457       & 14.3       \\ \hline
0.5                                                                      & 0.549        & 9.8        & \textbf{0.545}       & \textbf{9.0}       & \textbf{0.547}        & 9.4        & 0.558       & 11.6       & \textbf{0.450}        & \textbf{10.0}        & 0.568       & 13.6       & \textbf{0.539}        & \textbf{7.8}        & 0.566       & 13.2       \\ \hline
\end{tabular}
\label{tab:noisevar_gau}
\vspace{-10pt}
\end{table*}

\begin{table*}[t]
\centering
\caption{Noise Estimation Results in terms of $\hat{\sigma}$ and $\epsilon$ (\%) for Laplacian Noise}
\vspace{-10pt}
\begin{tabular}{|c|c;{0.5pt/1pt}c|c;{0.5pt/1pt}c|c;{0.5pt/1pt}c|c;{0.5pt/1pt}c|c;{0.5pt/1pt}c|c;{0.5pt/1pt}c|c;{0.5pt/1pt}c|c;{0.5pt/1pt}c|c|}
\hline
\multirow{3}{*}{\begin{tabular}[c]{@{}c@{}}Actual\\ $\sigma$\end{tabular}} & \multicolumn{16}{c|}{Estimated $\sigma$ ($\hat{\sigma}$) and $\epsilon$ (\%) respectively}                                                                                                                                                                             \\ \cline{2-17} 
                                                                          & \multicolumn{4}{c|}{Bunny}                            & \multicolumn{4}{c|}{Gargoyle}                         & \multicolumn{4}{c|}{DC}                               & \multicolumn{4}{c|}{Daratech}                         \\ \cline{2-17} 
                                                                          & \multicolumn{2}{c|}{Prop.} & \multicolumn{2}{c|}{PCA} & \multicolumn{2}{c|}{Prop.} & \multicolumn{2}{c|}{PCA} & \multicolumn{2}{c|}{Prop.} & \multicolumn{2}{c|}{PCA} & \multicolumn{2}{c|}{Prop.} & \multicolumn{2}{c|}{PCA} \\ \hline
0.1                                                                      & \textbf{0.117}        & \textbf{17.0}        & 0.123       & 23.0       & \textbf{0.112}        & \textbf{12.0}        & 0.119       & 19.0       & \textbf{0.120}        & \textbf{20.0}        & 0.124       & 24.0       & \textbf{0.118}        & \textbf{18.0}        & 0.124       & 24.0       \\ \hline

0.2                                                                      & \textbf{0.221}        & \textbf{10.5}        & 0.244       & 22.0       & \textbf{0.225}        & \textbf{12.5}        & 0.239       & 19.5       & \textbf{0.234}        & \textbf{17.0}        & 0.245       & 22.5       & \textbf{0.239}        & \textbf{19.5}        & 0.253       & 26.5       \\ \hline

0.3                                                                      & \textbf{0.332}        & \textbf{10.7}        & 0.372       & 24.0       & \textbf{0.333}        & \textbf{11.0}        & 0.387       & 29.0       & \textbf{0.337}        & \textbf{12.3}        & 0.384       & 28.0       & \textbf{0.344}        & \textbf{14.7}        & 0.384       & 28.0       \\ \hline

0.4                                                                      & \textbf{0.443}        & \textbf{10.8}        & 0.496       & 24.0       & \textbf{0.455}        & \textbf{13.8}        & 0.501       & 25.3       & \textbf{0.462}        & \textbf{15.5}        & 0.524       & 31.0       & \textbf{0.455}        & \textbf{13.7}        & 0.537       & 34.3       \\ \hline
0.5                                                                      & \textbf{0.546}        & \textbf{9.2}        & 0.587       & 17.4       & \textbf{0.558}        & \textbf{11.6}        & 0.601       & 20.2       & \textbf{0.571}        & \textbf{14.2}        & 0.627       & 25.4       & \textbf{0.564}        & \textbf{12.8}        & 0.639       & 27.8       \\ \hline
\end{tabular}
\label{tab:noisevar_lap}
\vspace{-10pt}
\end{table*}

\begin{table*}[t]
\centering
\caption{Denoising Accuracy for Gaussian Noise ($\sigma~=~0.2$) in terms of C2C and C2P ($\times 10^{-2}$) Respectively}
\vspace{-10pt}
\begin{tabular}{|c|c;{0.5pt/1pt}c|c;{0.5pt/1pt}c|c;{0.5pt/1pt}c|c;{0.5pt/1pt}c|c;{0.5pt/1pt}c|c;{0.5pt/1pt}c|c;{0.5pt/1pt}c|c;{0.5pt/1pt}c|}
\hline
\multirow{2}{*}{Model} & \multicolumn{2}{c|}{\multirow{2}{*}{Noise}} & \multicolumn{2}{c|}{\multirow{2}{*}{APSS}} & \multicolumn{2}{c|}{\multirow{2}{*}{RIMLS}} & \multicolumn{2}{c|}{\multirow{2}{*}{AWLOP}} & \multicolumn{2}{c|}{\multirow{2}{*}{MRPCA}} & \multicolumn{2}{c|}{\multirow{2}{*}{GTV}} & \multicolumn{4}{c|}{Proposed (Section~\ref{sec:sol for l1})}                      \\ \cline{14-17} 
                       & \multicolumn{2}{c|}{}                       & \multicolumn{2}{c|}{}                      & \multicolumn{2}{c|}{}                       & \multicolumn{2}{c|}{}                       & \multicolumn{2}{c|}{}                       & \multicolumn{2}{c|}{}                     & \multicolumn{2}{c|}{CG} & \multicolumn{2}{c|}{GSF} \\ \hline
Bunny                  & 0.266                & 2.81                & 0.209                & 0.474               & 0.219                & 0.891                & 0.262                & 2.49                & 0.222                & 1.108                & 0.202               & 0.483               & 0.193      & 0.467      & \textbf{0.191}       & \textbf{0.463}      \\ \hline
Gargoyle               & 0.248                & 2.45                & 0.194                & 0.634               & 0.204                & 0.995                & 0.238                & 2.13                & 0.199                & 0.836                & 0.188               & 0.711               & \textbf{0.173}      & 0.659      & 0.175       & \textbf{0.611}      \\ \hline
DC                     & 0.249                & 2.47                & 0.191                & 0.454               & 0.203                & 0.876                & 0.246                & 2.40                & 0.193                & 0.535                & 0.188               & 0.451               & \textbf{0.172}      & \textbf{0.428}      & 0.178       & 0.439      \\ \hline
Daratech               & 0.255                & 2.55                & 0.206                & 0.805               & 0.215                & 1.122                & 0.244                & 2.19                & 0.226                & 1.589                & 0.194               & 1.014               & 0.182      & 0.971      & \textbf{0.179}       & \textbf{0.937}      \\ \hline
Anchor                 & 0.258                & 2.67                & 0.202                & 0.532               & 0.207                & 0.678                & 0.245                & 2.06                & 0.196                & 0.275                & 0.191               & 0.201               & \textbf{0.177}      & \textbf{0.191}      & 0.178       & 0.195      \\ \hline
Lordquas               & 0.253                & 2.55                & 0.205                & 0.690               & 0.210                & 0.909                & 0.243                & 1.61                & 0.215                & 1.194                & 0.202               & 0.702               & 0.188      & 0.612      & \textbf{0.183}       & \textbf{0.607}      \\ \hline
Fandisk                & 0.296                & 3.35                & 0.279                & 2.426               & 0.259                & 1.454                & 0.287                & 2.88                & 0.248                & 1.201                & 0.214               & 1.137               & \textbf{0.194}      & 1.012      & 0.197       & \textbf{1.001}      \\ \hline
Laurana                & 0.250                & 2.50                & 0.190                & 0.390               & 0.203                & 0.849                & 0.238                & 2.07                & 0.190                & 0.374                & 0.183               & 0.384               & 0.174      & 0.348      & \textbf{0.169}       & \textbf{0.319}      \\ \hline
\end{tabular}
\label{tab:gauusian_0.2}
\vspace{-10pt}
\end{table*}

\begin{table*}[t]
\centering
\caption{Denoising Accuracy for Gaussian Noise ($\sigma~=~0.4$) in terms of C2C and C2P ($\times 10^{-2}$) Respectively}
\vspace{-10pt}
\begin{tabular}{|c|c;{0.5pt/1pt}c|c;{0.5pt/1pt}c|c;{0.5pt/1pt}c|c;{0.5pt/1pt}c|c;{0.5pt/1pt}c|c;{0.5pt/1pt}c|c;{0.5pt/1pt}c|c;{0.5pt/1pt}c|}
\hline
\multirow{2}{*}{Model} & \multicolumn{2}{c|}{\multirow{2}{*}{Noise}} & \multicolumn{2}{c|}{\multirow{2}{*}{APSS}} & \multicolumn{2}{c|}{\multirow{2}{*}{RIMLS}} & \multicolumn{2}{c|}{\multirow{2}{*}{AWLOP}} & \multicolumn{2}{c|}{\multirow{2}{*}{MRPCA}} & \multicolumn{2}{c|}{\multirow{2}{*}{GTV}} & \multicolumn{4}{c|}{Proposed (Section~\ref{sec:sol for l1})}                      \\ \cline{14-17} 
                       & \multicolumn{2}{c|}{}                       & \multicolumn{2}{c|}{}                      & \multicolumn{2}{c|}{}                       & \multicolumn{2}{c|}{}                       & \multicolumn{2}{c|}{}                       & \multicolumn{2}{c|}{}                     & \multicolumn{2}{c|}{CG} & \multicolumn{2}{c|}{GSF} \\ \hline
Bunny                  & 0.377                & 6.23                & 0.254                & 1.66               & 0.273                & 2.59                & 0.360                & 5.14                & 0.243                & 1.420                & 0.244               & 1.435               & \textbf{0.214}      & \textbf{1.178}      & 0.217       & 1.207      \\ \hline
Gargoyle               & 0.352                & 5.36                & 0.242                & 2.12               & 0.258                & 2.96                & 0.333                & 4.67                & 0.236                & 1.770                & 0.232               & 1.832               & 0.223      & 1.610      & \textbf{0.218}       & \textbf{1.471}      \\ \hline
DC                     & 0.354                & 5.50                & 0.233                & 1.65               & 0.257                & 2.78                & 0.330                & 4.61                & 0.227                & 1.310                & 0.218               & 1.294               & 0.201      & 1.087      & \textbf{0.195}       & \textbf{1.035}      \\ \hline
Daratech               & 0.357                & 5.05                & 0.295                & 3.78               & 0.322                & 5.03                & 0.331                & 4.14                & 0.292                & 3.030                & 0.274               & 3.107               & 0.261      & 2.413      & \textbf{0.259}       & \textbf{2.139}      \\ \hline
Anchor                 & 0.367                & 5.82                & 0.244                & 1.60               & 0.249                & 1.84                & 0.326                & 4.07                & 0.228                & 0.760                & 0.215               & 0.251               & \textbf{0.205}      & \textbf{0.238}      & 0.208       & 0.245      \\ \hline
Lordquas               & 0.372                & 5.97                & 0.262                & 1.71               & 0.280                & 2.58                & 0.284                & 2.44                & 0.258                & 1.580                & 0.247               & 1.321               & \textbf{0.224}      & \textbf{1.271}      & 0.228       & 1.295      \\ \hline
Fandisk                & 0.495                & 9.69                & 0.404                & 3.76               & 0.401                & 3.57                & 0.473                & 8.35                & 0.378                & 1.870                & 0.351               & 1.935               & 0.329      & \textbf{1.628}      & \textbf{0.323}       & 1.647      \\ \hline
Laurana                & 0.356                & 5.58                & 0.229                & 1.45               & 0.246                & 2.21                & 0.332                & 4.58                & 0.220                & 0.933                & 0.207               & 0.971               & \textbf{0.187}      & \textbf{0.781}      & 0.192       & 0.815      \\ \hline
\end{tabular}
\label{tab:gauusian_0.4}
\vspace{-12pt}
\end{table*}

\begin{table*}[t]
\centering
\caption{Denoising Accuracy for Laplacian Noise ($\sigma~=~0.1$) in terms of C2C and C2P ($\times 10^{-3}$) Respectively}
\vspace{-10pt}
\begin{tabular}{|c|c;{0.5pt/1pt}c|c;{0.5pt/1pt}c|c;{0.5pt/1pt}c|c;{0.5pt/1pt}c|c;{0.5pt/1pt}c|c;{0.5pt/1pt}c|c;{0.5pt/1pt}c|c;{0.5pt/1pt}c|}
\hline
\multirow{2}{*}{Model} & \multicolumn{2}{c|}{\multirow{2}{*}{Noise}} & \multicolumn{2}{c|}{\multirow{2}{*}{APSS}} & \multicolumn{2}{c|}{\multirow{2}{*}{RIMLS}} & \multicolumn{2}{c|}{\multirow{2}{*}{AWLOP}} & \multicolumn{2}{c|}{\multirow{2}{*}{MRPCA}} & \multicolumn{2}{c|}{\multirow{2}{*}{GTV}} & \multicolumn{2}{c|}{\multirow{2}{*}{\begin{tabular}[c]{@{}c@{}}Proposed\\ (Section~\ref{sec:sol for l1})\end{tabular}}} & \multicolumn{2}{c|}{\multirow{2}{*}{\begin{tabular}[c]{@{}c@{}}Proposed\\ (Section~\ref{sec:l2fid})\end{tabular}}} \\
                       & \multicolumn{2}{c|}{}                       & \multicolumn{2}{c|}{}                      & \multicolumn{2}{c|}{}                       & \multicolumn{2}{c|}{}                       & \multicolumn{2}{c|}{}                       & \multicolumn{2}{c|}{}                     & \multicolumn{2}{c|}{}  & \multicolumn{2}{c|}{}                         \\ \hline
Bunny                  & 0.144                & 9.24                 & 0.124                & 2.60                & 0.134                & 5.07                 & 0.140                & 7.38                 & 0.129                & 3.88                 & 0.119               & 2.18               & 0.114                  & 1.95 & \textbf{0.101} & \textbf{1.72}                 \\ \hline
Gargoyle               & 0.141                & 8.53                 & 0.124                & 3.21                & 0.134                & 5.87                 & 0.137                & 7.50                 & 0.136                & 6.41                 & 0.121               & 4.57               & 0.116                  & 3.16 & \textbf{0.104} & \textbf{2.63}                 \\ \hline
DC                     & 0.141                & 8.66                 & 0.120                & 2.30                & 0.130                & 4.65                 & 0.136                & 6.97                 & 0.126                & 3.77                 & 0.117               & 2.28               & 0.113                  & 2.19 & \textbf{0.098} & \textbf{1.67}                 \\ \hline
Daratech               & 0.142                & 8.87                 & 0.124                & 3.03                & 0.127                & 3.68                 & 0.142                & 8.72                 & 0.140                & 7.28                 & 0.120               & 3.17               & 0.113                  & 3.04 & \textbf{0.101} & \textbf{2.38}                 \\ \hline
Anchor                 & 0.142               & 8.97                & 0.123                & 3.30               & 0.128                & 3.96                & 0.139                & 7.64                & 0.119                & 1.71                & 0.121               & 3.11               & 0.116                  & 1.95 & \textbf{0.104} & \textbf{1.32}                 \\ \hline
Lordquas               & 0.137                & 8.23                 & 0.125                & 4.05                & 0.131                & 5.48                 & 0.135                & 7.48                 & 0.124                & 3.90                 & 0.119               & 3.80               & 0.113                  & 3.74 & \textbf{0.102} & \textbf{2.97}                 \\ \hline
Fandisk                & 0.145                & 9.15                 & 0.138                & 6.83                & 0.138                & 6.91                 & 0.144                & 8.27                 & 0.127                & 3.15                 & 0.131               & 4.27               & 0.126                  & 4.17 & \textbf{0.110} & \textbf{2.53}                 \\ \hline
Laurana                & 0.141                & 8.79                 & 0.120                & 2.40                & 0.133                & 5.62                 & 0.139                & 8.18                 & 0.121                & 2.69                 & 0.120               & 2.97               & 0.117                  & 2.61 & \textbf{0.105} & \textbf{2.07}                 \\ \hline
\end{tabular}
\label{tab:laplace_0.1}
\vspace{-10pt}
\end{table*}

\begin{table*}[t]
\centering
\caption{Denoising Accuracy for Laplacian Noise ($\sigma~=~0.3$) in terms of C2C and C2P ($\times 10^{-2}$) Respectively}
\vspace{-10pt}
\begin{tabular}{|c|c;{0.5pt/1pt}c|c;{0.5pt/1pt}c|c;{0.5pt/1pt}c|c;{0.5pt/1pt}c|c;{0.5pt/1pt}c|c;{0.5pt/1pt}c|c;{0.5pt/1pt}c|c;{0.5pt/1pt}c|}
\hline
\multirow{2}{*}{Model} & \multicolumn{2}{c|}{\multirow{2}{*}{Noise}} & \multicolumn{2}{c|}{\multirow{2}{*}{APSS}} & \multicolumn{2}{c|}{\multirow{2}{*}{RIMLS}} & \multicolumn{2}{c|}{\multirow{2}{*}{AWLOP}} & \multicolumn{2}{c|}{\multirow{2}{*}{MRPCA}} & \multicolumn{2}{c|}{\multirow{2}{*}{GTV}} & \multicolumn{2}{c|}{\multirow{2}{*}{\begin{tabular}[c]{@{}c@{}}Proposed\\ (Section~\ref{sec:sol for l1})\end{tabular}}} & \multicolumn{2}{c|}{\multirow{2}{*}{\begin{tabular}[c]{@{}c@{}}Proposed\\ (Section~\ref{sec:l2fid})\end{tabular}}} \\
                       & \multicolumn{2}{c|}{}                       & \multicolumn{2}{c|}{}                      & \multicolumn{2}{c|}{}                       & \multicolumn{2}{c|}{}                       & \multicolumn{2}{c|}{}                       & \multicolumn{2}{c|}{}                     & \multicolumn{2}{c|}{}  & \multicolumn{2}{c|}{}                           \\ \hline
Bunny                  & 0.303                & 3.82                & 0.221                & 1.183               & 0.235                & 1.39                & 0.294                & 3.17                & 0.230                & 1.249                & 0.221               & 1.37               & 0.214                  & 1.258 & \textbf{0.193} & \textbf{0.875}                 \\ \hline
Gargoyle               & 0.284                & 3.30                & 0.208                & 1.199               & 0.221                & 1.61                & 0.270                & 2.91                & 0.208                & 1.103                & 0.205               & 1.20               & 0.197                  & 1.083 & \textbf{0.173} & \textbf{0.911}                 \\ \hline
DC                     & 0.285                & 3.30                & 0.203                & 1.177               & 0.219                & 1.41                & 0.233                & 1.79                & 0.202                & 1.020                & 0.194               & 1.01               & 0.186                  & 0.981 & \textbf{0.153} & \textbf{0.704}                 \\ \hline
Daratech               & 0.287                & 3.17                & 0.250                & 1.832               & 0.243                & 2.06                & 0.272                & 2.71                & 0.234                & 1.622                & 0.227               & 1.71               & 0.221                  & 1.688 & \textbf{0.194} & \textbf{1.237}                 \\ \hline
Anchor                 & 0.294                & 3.54                & 0.212                & 1.820               & 0.217                & 1.96                & 0.266                & 2.38                & 0.204                & 1.643                & 0.207               & 1.83               & 0.195                  & 1.753 & \textbf{0.168} & \textbf{1.325}                 \\ \hline
Lordquas               & 0.295                & 3.60                & 0.224                & 1.206               & 0.234                & 1.41                & 0.258                & 1.84                & 0.230                & 1.300                & 0.224               & 1.19               & 0.221                  & 1.043 & \textbf{0.208} & \textbf{1.018}                 \\ \hline
Fandisk                & 0.368                & 5.46                & 0.327                & 2.854               & 0.316                & 2.29                & 0.354                & 4.74                & 0.302                & 2.059                & 0.301               & 2.15               & 0.288                  & 2.075 & \textbf{0.259} & \textbf{1.694}                 \\ \hline
Laurana                & 0.287                & 3.38                & 0.200                & 0.648               & 0.214                & 1.18                & 0.268                & 2.70                & 0.198                & 0.549                & 0.201               & 1.13               & 0.192                  & 0.524 & \textbf{0.161} & \textbf{0.407}                 \\ \hline
\end{tabular}
\label{tab:laplace_0.3}
\vspace{-5pt}
\end{table*}

%% file: conclude.tex
We address the 3D point cloud denoising problem, where the fidelity term can be either $\ell_2$- or $\ell_1$-norm to accommodate different kinds of acquisition noise.
We propose a reweighted graph Laplacian regularizer (RGLR) for the surface normals as a signal prior with desirable properties, namely rotation-invariant and promotion of piecewise smoothness. 
To obtain a linear relationship between 3D coordinates and surface normals, we first perform bipartite graph approximation. 
For $\ell_2$-norm fidelity term, we iteratively solve a quadratic programming problem using conjugate gradient.
For $\ell_1$-norm fidelity term, we minimize the $\ell_1$-$\ell_2$-norm cost function using accelerated proximal gradient (APG). 
We derive computation-efficient mean and media filters to estimate noise variance for different noise types.
Experimental results show that our algorithms outperform competing schemes visually and quantitatively.